\documentclass{article}

\usepackage{arxiv}
\usepackage[utf8]{inputenc} 
\usepackage[T1]{fontenc}    
\usepackage{hyperref}       
\usepackage{url}            
\usepackage{booktabs}       
\usepackage{amsfonts}       
\usepackage{nicefrac}       
\usepackage{microtype}      
\usepackage{lipsum}
\usepackage{import}

\usepackage[
  backend=biber,
  style=numeric,
  sorting=none,
  maxnames=3
]{biblatex}
\bibliography{references}

\usepackage{graphicx}
\graphicspath{ {./figures/} }

\begin{document}

\title{Theoretical Model and Practical Considerations for Data Lineage Reconstruction}
\author{
  Egor Pushkin\\
  AWS AI\\
  \texttt{ypushki@amazon.com} \\
}
\maketitle

\begin{abstract}

We live in a world driven by data. The amount of it outgrows anyone's ability to oversee it or even observe its scope. Along with all the advances in the space of data management, there is still a significant lack of formalism and standardization around defining data ecosystems and processes occurring within those. In order to address the issue we propose a notation for data flow modeling and evaluate some of the most common applications of it based on real-world use cases. To facilitate future work, we provide detailed reference of the data model we defined and consider potential programming paradigms.

\end{abstract}

\section{Introduction}
\label{sec:intro}

Variety of data ecosystems grow in all areas of information technology. The amount of data collected on a daily basis increases exponentially fueled by advances in edge collection techniques (IoT) and our ability to store and process vast amounts of it\footnote{It is expected to reach the volume of 175 zettabytes by as early as 2025 \cite{sea_data_age} with approximate daily production of 463 exabytes \cite{rac_day_data}.}. Rapid grows of it gave birth to a variety of platforms targeted at those areas and as a result led to technical and organizational fragmentation. There is no shortage of ways data is being produced, transformed, and stored \cite{def_arc_big_data} even within closed boundaries of individual organizations. 

We already reached the state where no job (manual or automated) is done is isolation. No person (or a program) starts execution of a task sitting at an empty desk with a blank page and a pen in front of them. Every activity involves consumption, transformation of data, and production of derivatives of it. Individuals and organizations structure and redefine their workflows in a data-driven manner. 

All that in mind, most data manipulations remain almost completely untraceable \cite{data_prov_basic_iss}. Even most sophisticated infrastructures are lacking basic abilities for the analysis of data routes and ways data flows through it. This state of things leads to oversights in data flow design, introduces compliance violations \cite{root_cause_compl_vl} and opens the doors for unintended access to sensitive data \cite{top_sec_threats}. 

The importance of addressing this issue rises along with growing dependence on data systems. This work proposes a theoretical model of describing flow of data with recoverable lineage. The goal of this research is to lay down the framework for formal representation of data processing system, practically speaking, any system we operate today. 

We position this proposal as a specification for a standalone system that can be used for tracking data manipulations in technology- and platform-agnostic manner. The goal of this proposal is to be independent of any specific implementation of data transformation and data storage paradigm. 

Among all of the use cases we aim to enable, the following ones are worth highlighting as the most essential: 
\begin{itemize}
\item Ability to record the state of data system to improve visibility;
\item Enable reproducibility by formally describing every transformation data undergoes;
\item Improve auditability of the system;
\item Provide compliance conformance monitoring and enforcement; 
\item Honor privacy of data contributors; 
\item Enforce stronger levels of security and simplify gap analysis. 
\end{itemize}

Here and below we refer to the data as any piece of information that can be stored in a digital form. The method we document scales even beyond digital world but we ignore that for consistency. We do not, however, limit or detail the type, format or modality of data the system operates with. 

The paper is structured as follows: We start with an overview of related work previously performed in the field. In Section \ref{sec:model}, we explain major principles of proposed paradigm for data lineage tracking. Section \ref{sec:scaling} briefly covers challenges and potential solutions for scaling the model to large environment. We then proceed to a deeper review of a number of practical considerations around the adoption of the model in Section \ref{sec:examples}. Section \ref{sec:conclusion} summarizes our conclusions and forward looking vision. Finally, Appendix \ref{sec:progr} provides a glimpse into programming model and Appendix \ref{sec:reference} contains detailed references of all entities being discussed in the paper. 

\section{Related Work}
\label{sec:related}

While to the best of our knowledge there is no direct equivalent to the system described in our proposal we review a number of areas that are adjacent in its meaning or functionality. In a sense any paradigm or platform targeted at dealing with data at scale can be considered related to this work in one way or another. 

Coincidentally the closest segment of tools to our work is found in a relatively recently formed domain of machine learning life cycle and experiment tracking. Some of the notable examples in the space are mlflow \cite{rel_mlflow}, comet.ml \cite{rel_comet_ml}, CodaLab \cite{rel_codalab}. While a certain level of similarity can be observed in there, those services are over indexing on a specific niche area and are lacking formalism and usability required for wider adoption. 

Due to being data-driven by nature this proposal overlaps with data processing and distribution frameworks like Apache NiFi \cite{rel_nifi}. It provides great level of visibility into data flow through its concept of provenance store. However it deviates from our goals in a number of key elements being very prescriptive and opinionated regarding execution environment and heavy focused towards streaming use cases. 

Another important component of data universe worth mentioning is family of metadata exploration services (cloud: AWS Glue Data Catalog \cite{rel_aws_glue_dc}, Azure Data Catalog \cite{rel_azure_dc}; open source: Apache Atlas \cite{rel_atlas}, Netflix MetaCat \cite{rel_metacat}; proprietary: Uber Databook \cite{rel_databook}). These offerings have a lot in common and primarily focus on identifying various locations where data lives. Being extremely helpful for discovery they rarely go into the level of details needed to track individual pieces (or slices) of data stored in respective locations. 

While some tools and platforms are emerging in the space most data engineers still find it hard to adopt it for daily tasks and high-pace experimentation. This state of things defines the problem as unsolved and appeals for the additional research in this direction to be performed and a set of acceptable and widely available solutions to emerge from it. 

The overview of existing data storage, source code management, execution environments, and schema registries is omitted primarily because the work presented here while refers extensively does not have any overlap with those systems in terms of functionality and does not aim at replacing any of its features. 

\section{Data Model}
\label{sec:model}

The purpose of this proposal is to define formal model suitable for mapping data manipulations occurring in complex environments. We do not aim at covering specific process or targeting a particular technology stack. Concrete examples we provide along the way are mostly used for clarification purposes. 

On the contrary, the goal of this work is to be able to scale to a wide range of potential applications. We keep OCP \cite{obj_orient_soft_constr} in mind when laying out fundamentals and data model skeleton (covered in full in Appendix \ref{sec:reference}). We do not anticipate the model described here to be complete but rather aim at establishing the right foundation that can be further extended and built upon. 

In this section we provide an overview of major building blocks and principles of the model. We cover notations, semantics and primitives required for rendering comprehensive picture of data flow process. 

\subsection{Fundamentals}

Data lineage system \cite{data_lin_survey,data_lin_prov_es,pr_data_lin_wh,fin_data_lin_db} is defined through its ability to say what happens with any piece of data registered on the system at any point of time and establish source-target relationships in between those. 

In this framework, we draw the line in between data lineage system and other components of software development life cycle and live application as it is defined and deployed. The key principle is formulated in terms of separation of responsibilities between the above-mentioned actors. Data lineage model defined here is used exclusively for tracking purposes and does not extend to areas covered by existing software systems and development tools. In practice this specifically means that data lineage system does not provide neither storage and execution capabilities. 

Such approach is driven by the following reasoning and requirements: 

\begin{itemize}
\item Lineage platform should not be dependent on any components of existing applications. 
\item Referential data maintained by the system should not contain customer or sensitive data. This requirement is driven by the desire to be able to preserve data manipulations log independently of data life cycle defined by external compliances \cite{compl_gdpr,compl_hipaa} and internal procedures. 
\item Being positioned as a standalone system offers great flexibility in terms of design and usage patterns. 
\item Being deployed separately provides critical level of independence from existing live applications that extends in both directions (protects both sides). 
\item Strict separation allows data lineage environment to provide tracking coverage for a wide set of diverse data and execution platforms that are incompatible by nature and are unaware of each other's existence even when are used in combination as a part of the same data processing flow. 
\item Data lineage system should be easily pluggable into existing existing and new heterogeneous environments. This is achieved by putting no requirements on application design and implementation. It is essential to ensure that the integration of tracking platform can occur at code level without application redesign. 
\end{itemize}

In the scope of this proposal data lineage system is seen as activity recording platform providing extensive support for analysis. The framework described here defines finite system. All activities tracked by it occur within boundaries defined by host application. Host application controls the way data is being managed outside of the lineage model and draws the line of how that boundary is being defined. 

Cornerstone element of proposed lineage model is a concept of data mutation or transformation that converts a data from one representation to another:

\begin{equation}
D' = T(D)
\end{equation}

which can be extended to indicate that transforms deal with multiple data blobs and both sides undergo continuous iteration: 

\begin{equation}
[D':R', ...] = T([D:R, ...])
\end{equation}

Trivial machine learning flow can be represented in the following way using this semantics: 

\begin{equation}
Model, EvalMetrics = Eval(Train(PreTrained, PreProc(TrainSet)), TestSet)
\end{equation}

The model we further describe and evaluate in this proposal focuses on enabling the following capabilities that can be combined to power complex analysis:

\begin{itemize}
\item \textbf{Forward Tracing}\enspace The operation produces complete list of data products (with references to involved transform executions) derived off a specific piece of data. It provides ability to see how data products are being spread across the system and observe areas impacted by those. Identified dependencies can be used for passive analysis as well as active broadcasting of compliance notifications. 
\item \textbf{Backward Tracing}\enspace Reverse lookup initiated from a data blob leading to the entire hierarchy of its direct ascendants. Lookups of such sort prove itself especially useful when analyzing the impact of data and code changes occurring over time. Software systems consisting of various interconnected components present a great challenge for this type of analysis when investigation crosses the boundaries of components.
\item \textbf{Environment Definition}\enspace Can be considered being an extension of backward tracing and defines the state of the environment (a combination of data and performed transformations) required for complete reproduction of a specific dataset. 
\item \textbf{System Evolution}\enspace Provides an additional dimension to traversal capabilities allowing data relations to be tied to iterations the system being monitored went through in chronological order. 
\item \textbf{Actor Participation}\enspace Looks at the system from the perspective of actors responsible for all data manipulations performed. Actor information combines awareness of parties involved in the process of data flow with reasoning and additional context around performed mutations. 
\end{itemize}

In order to achieve these objectives we define a framework suited for mapping data flows to the set of basic primitives covering both data products and manipulations. The essential component of this approach is the level of flexibility it offers. It provides recommendations and only resorts to rules in the areas where those are essential for achieving key goals and avoiding design ambiguity. The elements of the toolbox are picked to simplify system definition in conformance with best practices and paradigms of building data-driven applications. 

\subsection{Notation}

\begin{figure}[htbp]
  \centering
  \includegraphics[scale=0.6]{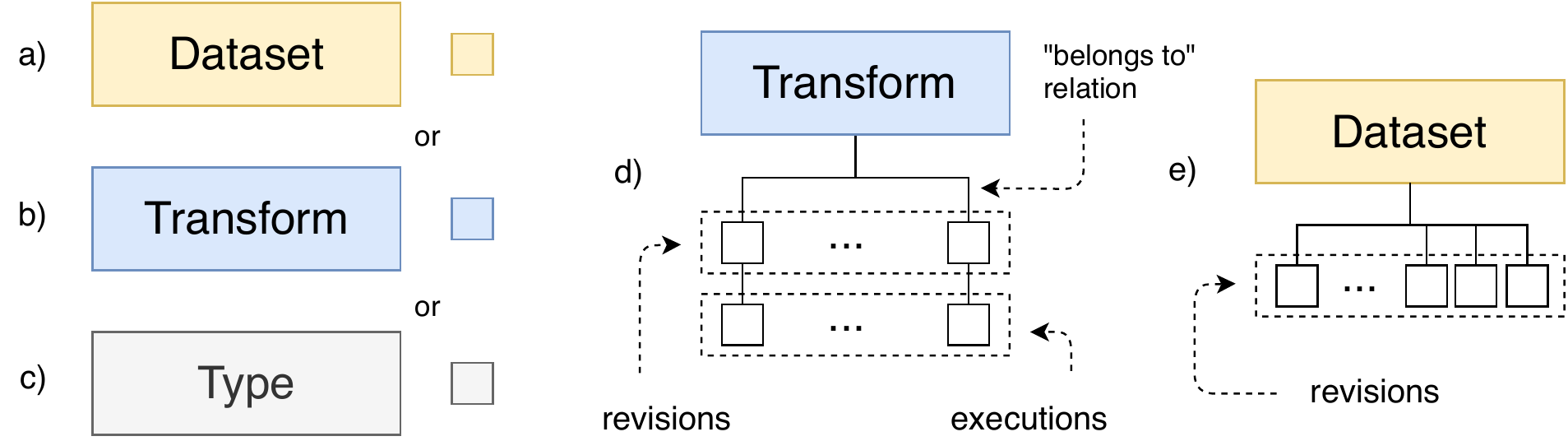}
  \caption{Notations used in drawings throughout the paper: a) standalone dataset object; b) transform object; c) type object; d) transform object with revisions (layer 1) and executions (layer 2); e) dataset object with its revisions.}
  \label{fig:not_graph}
\end{figure}

Illustrations and figures provided in the paper use consistent notation for visualizing major entities being discussed. The notation is explained in Figure \ref{fig:not_graph}.

Table \ref{table:not_ents} mentions common notations present in textual form. 

\begin{table}[!htbp]
  \centering
  \begin{tabular}{p{0.1\linewidth}p{0.5\linewidth}p{0.25\linewidth}}
    \toprule
    Elements    & Definition        & Examples \\
    \midrule
    Entity Id    & \texttt{[entity] \_ [identifier]} & \(DS_{in}\) \\
    Entity       & \texttt{[dataset] ":" [revision]} & \(DS_i:R_j\) \\
                 & \texttt{[transform] ":" [revision] ":" [execution]} & \(TR_i:R_j:E_k\) \\
    Route        & \texttt{"[" [entity] ("," ...) "]"} & \([DS_1, DS_2]\) \\ 
    List         & \texttt{"L(" [entity or route] ("," ...) ")"} & \(L(DS_i:R_j, [DS_1, DS_2])\) \\
    \bottomrule
  \end{tabular}
  \caption{Textual representation of entities.}
  \label{table:not_ents}
\end{table}

Object identifiers play an important role in the mechanics of the proposal. We do not detail the mechanism used for generating those but emphasize that identifiers must remain unique within the scope of a particular entity type (dataset, dataset revision, transform). 

\begin{figure}[htbp]
  \centering
  \includegraphics[scale=0.6]{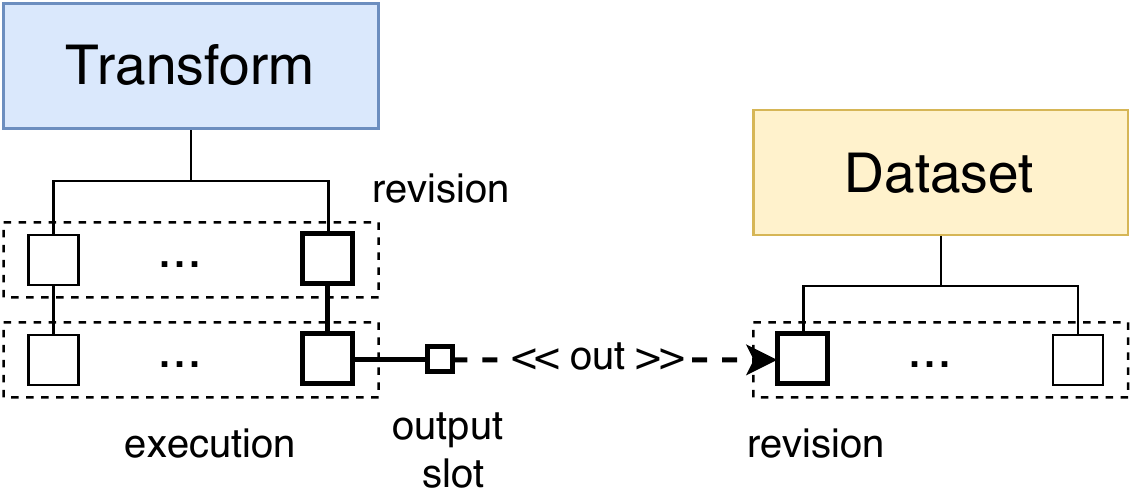}
  \caption{Basic mechanism of reference tracking between registered objects.}
  \label{fig:not_ids}
\end{figure}

Most diagrams omit the fact that all entity relations are established through its respective identifiers (see Figure \ref{fig:not_ids} for an example). Detailed reference of all entities and its relations discussed in the paper can be found in Appendix \ref{sec:reference}. 

\subsection{Datasets}
\label{sec:datasets}

In this work we take data-centric view of the world. We look at any system from the point of view of how data flows through it and aim at defining a formal model for documenting essential elements of propagation semantics. 

The approach we propose is based on the concept of a \textbf{dataset}. Dataset represents a piece of data tracked by the system. Data iterations performed by systems we analyzed and explored have one important characteristic in common: most transformations performed over data that is naturally bound (finite and unambiguously defined). Thus a concept of static snapshot deems to be sufficient for capturing the state and progress of any data flow. We use dataset as such abstraction. 

We do not formally define rules for the appropriate granularity level applications should adopt when designing their data flows using this toolbox (mapping to datasets, transforms, etc.). We indicate that the purpose of the semantics introduced here is to track and analyze changes in the data and therefore every individual piece of it that can ever be produced or versioned separately should be represented by an independent dataset instance. 

\subsubsection{Revisions}

Data naturally evolves over time within and outside of the digital world. The process of capturing changes in the ever evolving environment is a challenging task that deserves special attention in the scope of this work. In our proposal, dataset does not refer to the data itself but much rather acts as a container for a set of iterations of the same logical data blob produced over time. This could refer to data being regenerated as a consequence of changes made to input parameters of upstream process or naturally growing with new pieces being added. Each such iteration of a dataset is represented in the form of \textbf{dataset revision}. 

Dataset revisions as those are recorded on the lineage system refer to immutable data blobs. This fundamental principle comes with two significant consequences (in a sense those are also drivers of it). Every new piece of data recorded on the system is always recorded in the form of new dataset revision. Alongside with that, the contents of existing revisions (data referred by those) never changes. 

Revisions within dataset can be referred by its identifier or relative position (\(:earliest, :latest, :head-1\)). This provides an sufficient level of flexibility when defining transforms. Downstream applications can choose to implement additional semantics for querying revisions of dataset to take advantage of type information and custom metadata. 

\subsubsection{Storage}

One of the premises made in this proposal is that lineage system itself is abstracted from data, code storage and computational resources required for executing data transformations. The model described here is focused on keeping track of references to externally managed resources and identifying key relationships between those. This claim makes the system generic and applicable to a wide range of potential applications and flexible to operate independently from newly constructed and existing legacy systems. 

\begin{figure}[htbp]
  \centering
  \includegraphics[scale=0.6]{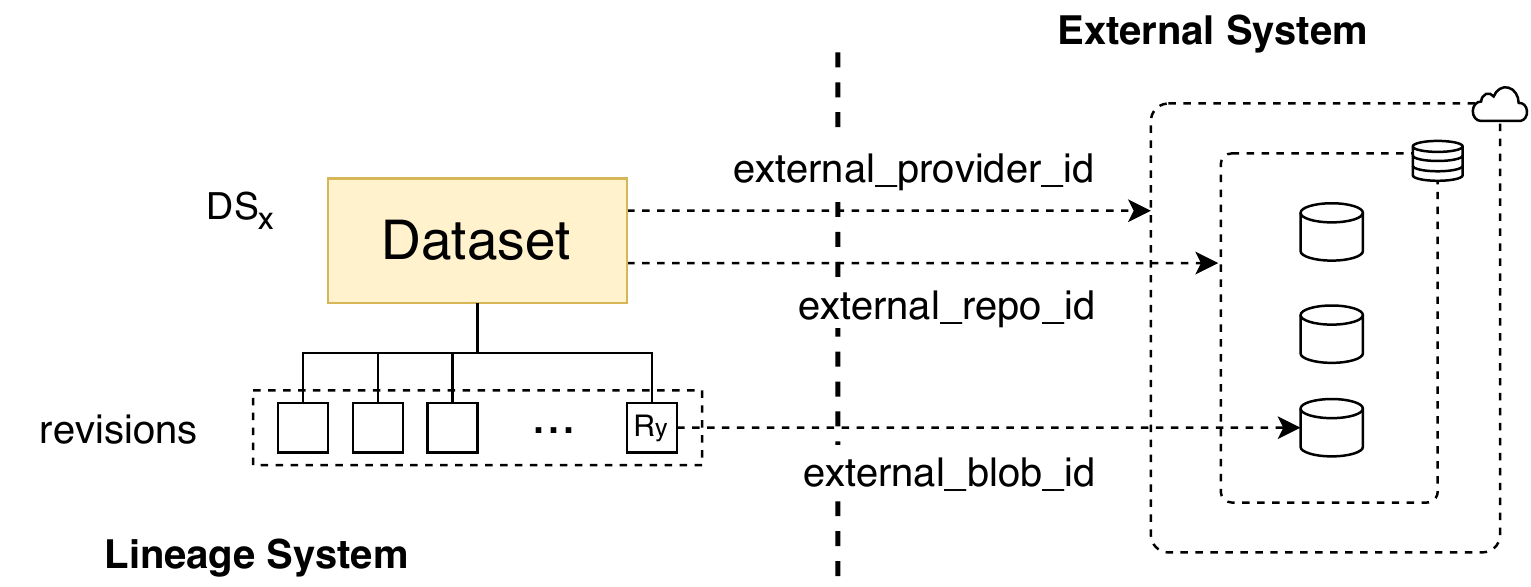}
  \caption{Relationship between lineage system and external data storage platform.}
  \label{fig:ds_stor}
\end{figure}

While being applicable to all entities tracked by the system, the concept of storage abstraction is especially important for data layer (see Figure \ref{fig:ds_stor}). A great number of digital data storage platforms are available out there in the industry \cite{ds_stor_pl_meth}. Being able to efficiently operate with all of those requires a well defined contract to be put in place. While deep integrations (that we will get back to in Appendix \ref{sec:progr}) usually take time (and may never happen for some closed systems) we need to have a solution in place that enables applications with proper level of tracking. This proposal focuses on the tier that is completely abstracted of any existing storage or execution platform. 

\subsubsection{Type System}

Data evolution and the process of registering new revisions is inevitably associated with the iteration on data formats (or data types) that host application natively understands and operates with. The model defined in the proposal allows data formats to be tracked and versioned through concepts of \textbf{type} and \textbf{type revision}. The pair of entities is introduced to logically complement datasets and dataset revisions. Type entity acts as a logical container of data formats represented by type revision. 

\begin{figure}[htbp]
  \centering
  \includegraphics[scale=0.6]{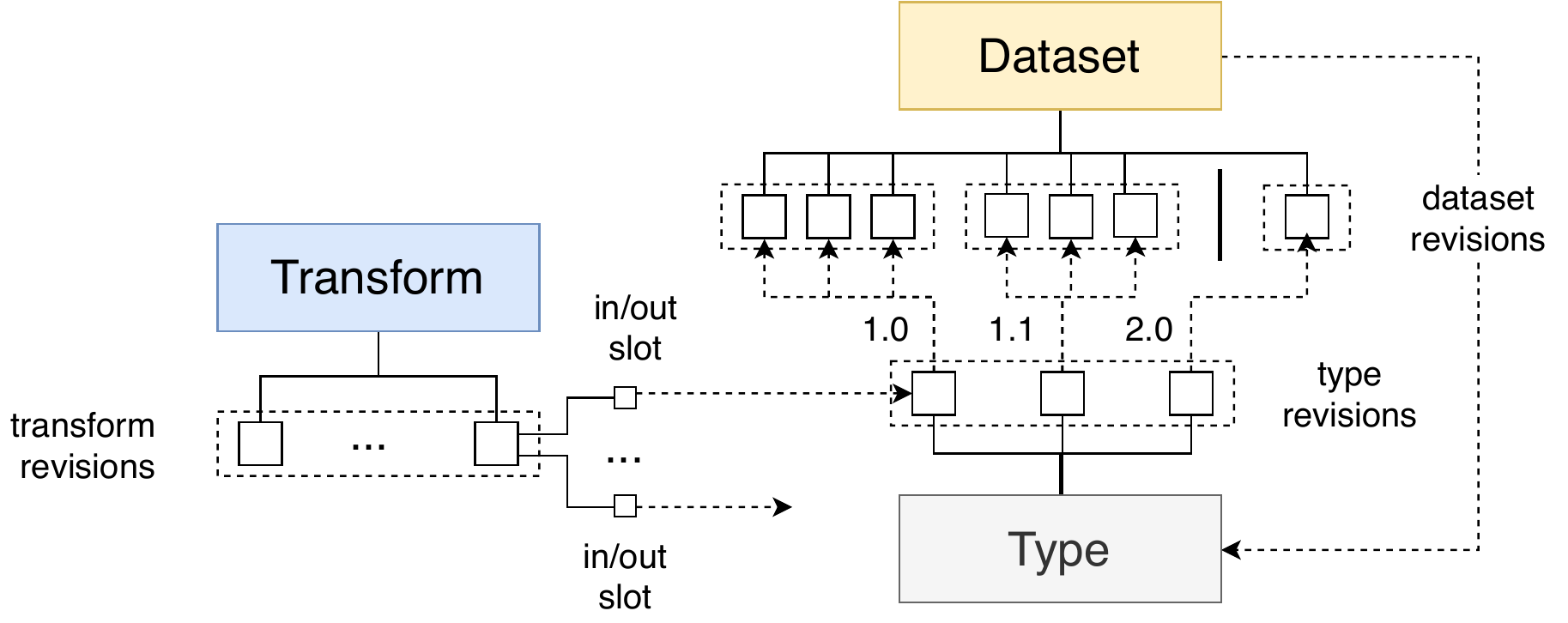}
  \caption{Relations between types, datasets and transform slots.}
  \label{fig:ds_types}
\end{figure}

Type revisions are reusable objects and could be referred to by multiple datasets. References to those (types) are propagated throughout the graph of dataset revisions automatically by being associated with input and output slots of respective data transforms (see Figure \ref{fig:ds_types}). Type information, as it propagates through the system, can be used by applications for in depth analysis of data and its evolution. 

While being able to record type information, data lineage system is built on the foundation of being unable to interpret it. Type definitions, that can be optionally associated with type revisions, are treated by the model as opaque blobs. This approach also manifests itself in the lack of subtyping available for datasets itself. Lineage model does not introduce subclasses of datasets narrowly focused on a specific data structure or a use case (key-value store, parameters bag, etc.). Such a level of detail is explicitly avoided to underline the separation defined between data lineage, data storage systems and host application tying those things together. On the other hand, applications equipped with type awareness can run queries against the model to determine and ensure compatibility of datasets and transforms, perform deeper analysis of the data, etc. 

\subsubsection{Streaming and Record-Level Data}

State of data stream is not being captured or managed directly but rather observed by data lineage system through slices (time, range or key-based) and/or compaction snapshots. Snapshot extraction and slicing is controlled by application business logic and is not strictly prescribed by the model. The model assumes that sufficient information regarding the status of the steam at the moment of snapshot extraction can be captured by the application in platform-specific manner and recorded in the snapshot dataset. A sequence of such point-in-time snapshots is represented in the form of dataset revisions. This information can be later used for lineage tracing beyond the boundaries of the framework described here. We cover details of such representation in Section \ref{sec:data_mod_tr}.

Stream processing logic described above is applicable to a wider range of input sources of data. Those include but are not limited to relational data, key-value stores, web content being scraped, user input accepted by upstream applications. 

\subsection{Transforms}
\label{sec:transforms}

\textbf{Transform} is the second fundamental building block of proposed model. Transform represents a piece of business logic interacting with zero or many datasets registered on the system.

\subsubsection{Revisions}

As the code evolves its changes are recorded in the form of \textbf{transform revisions}. Revision is logically equivalent to a commit made in external source version control system and points directly at it. Section \ref{sec:tracking_effect} contributes with details of how revisions can be leveraged for lineage analysis. 

Naturally transforms refer to external source version control system for storage of its revisions in the form of commits. In this work transform revisions are often treated as a list (vs. a tree intrinsic to most SVCs out there) which is sufficient for data lineage tracking purposes. Such separation provides ability to leverage existing infrastructure used for organizing software development life cycle.

\subsubsection{Executions}

\textbf{Transform execution} comes into play when transform is run. Executions incorporate runtime properties of transform revisions. While revisions provide static description of a transform, executions bind dataset revisions transform operates with. 

\subsubsection{Slots}

\begin{figure}[htbp]
  \centering
  \includegraphics[scale=0.6]{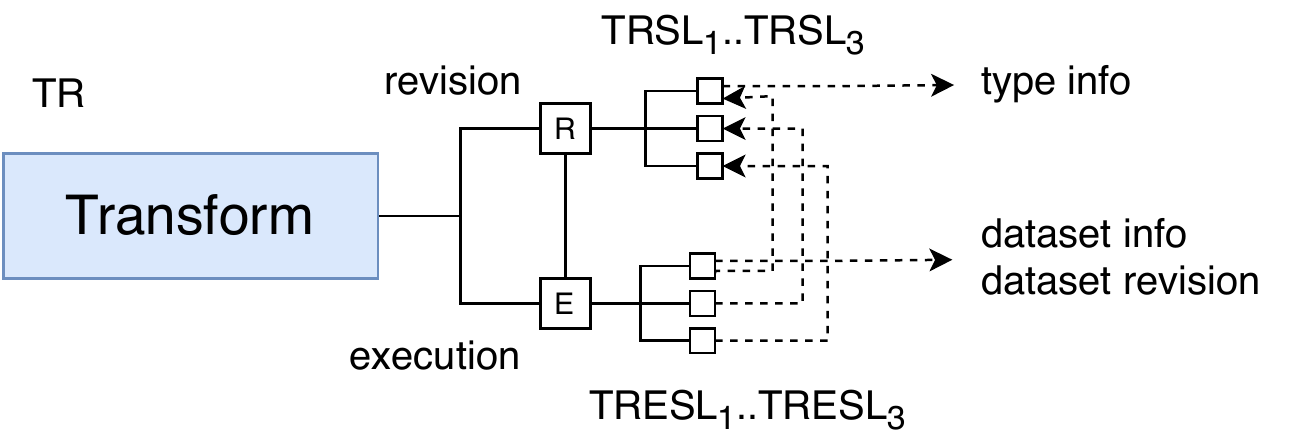}
  \caption{Transform revision, execution and its respective slots as it relates to each other.}
  \label{fig:tr_slots}
\end{figure}

Actual binding between transforms and datasets is implemented via an additional indirection level - \textbf{transform slots}. Main responsibility of slots is to identify the role a particular dataset revision plays within transform execution. Slots are defined and initialized with various properties at both revision and execution levels (see Figure \ref{fig:tr_slots}). Revision-level slots are typed (identify data format that revision expects to see as an input on that slot or produces as an output). Execution-level slots are filled in with a reference to specific dataset revision (produced or consumed). 

The fact that transform revision slots do not record any reference to a specific dataset/revision part is a clear indication of the principle that proposed model is focused on the ability to record changes occurring in the extraneous runtime environment and neither predicts nor dictates the behavior of external live system.

\subsubsection{Tracing Properties}

Tracing properties of transforms (or transform revisions) provide dimensions all transforms can be evaluated on for the purpose of better understanding of the rules of data propagation throughout the system. Tracing properties define behavioral aspects of transforms providing high-level view into its internals that can be achieved without deep analysis of implementation details (which is not always feasible). 

We evaluate and number of properties for demonstration purposes and as a guidance for further expansion:

\begin{itemize}
\item \textbf{Deterministic} Among numerous reasons for non-deterministic behavior \cite{on_determ_non_determ} the following ones are the most commonly faced with in practice: random generators (controlled via programmatically set seed), dependent libraries, hardware (e.g. GPUs). These factors can contribute to the execution of the same transform revision with the same input producing different output data. As an added bonus transform developers can choose to identify the level of non-determinism based on the significance of variance that can occur among transform executions. The impact of non-determinism can range from statistically insignificant to completely unpredictable. The understanding of this property plays an essential role in projection and analysis of reproducibility. 
\item \textbf{Reversible} Transforms possessing this property perform data manipulation that is completely or partially reversible meaning that output data products can be used to reproduce, simulate or otherwise expose the nature of input datasets. Examples of non-reversible transformations include various hashing algorithms, data profiling, metrics calculation, etc. 
\item \textbf{Privacy-preserving} It becomes increasingly important for applications to honor and properly manage privacy of various actors entrusting it to store and manage their data. Variety of techniques have been developed in this direction and can be used as flags under this general property. Some of those are: anonymization \cite{priv_anon_tech}, personal information reduction (PII, PCI, PHI, etc.) \cite{priv_pii_red_book}, differential privacy \cite{priv_diff_survey}. 
\item \textbf{Generative} Generative transforms produce new data (based on its input or even with the lack thereof). In the current definition of the property, data produced by these transforms does not contain elements of input datasets and is generated by the rules described at code level. Generative transforms can, however, produce data based on templates or some other definitions provided externally. 
\end{itemize}

In its most trivial definition these properties (and others introduced by downstream applications) can be leveraged when implementing traceability algorithms for answering questions whether tracing should continue through the specific transform or it can be considered terminal point of traversal (see Section \ref{sec:track_fund}). 

It is important to underline that properties discussed here are most often manually specified and are not easily provable. It is the responsibility of transform developers to summarize its behavior in such manner and indicate ambiguity where it exists. Some of these properties exist in the grey areas of determinism and reversibility and in combination with lack of formal provability present challenge for unconditional reliance upon them. That said, it can still be used as an important optimization and approximation mechanism for evaluation of data processes in large data ecosystems that are not easily understood otherwise. 

As a safety measure applications can choose to implement regression tests and runtime monitoring systems that are continuously tracking the validity of transform properties. Even though most of these properties are not formally provable, some can be validated empirically. 

\subsubsection{Flows}

It is rare that data processing flows consist of individual transforms. More often than not data goes through a number of processing steps until it reaches its final form. The sequence of those executions and relationships between resulting data products present high value for subsequent analysis. Therefore we introduce the concept of \textbf{flow} to capture static structure of multi-step data flow, corresponding \textbf{flow revision} for tracking changes in such a formation, and \textbf{flow execution} to record dynamic relations established at run time. 

\begin{figure}[htbp]
  \centering
  \includegraphics[scale=0.6]{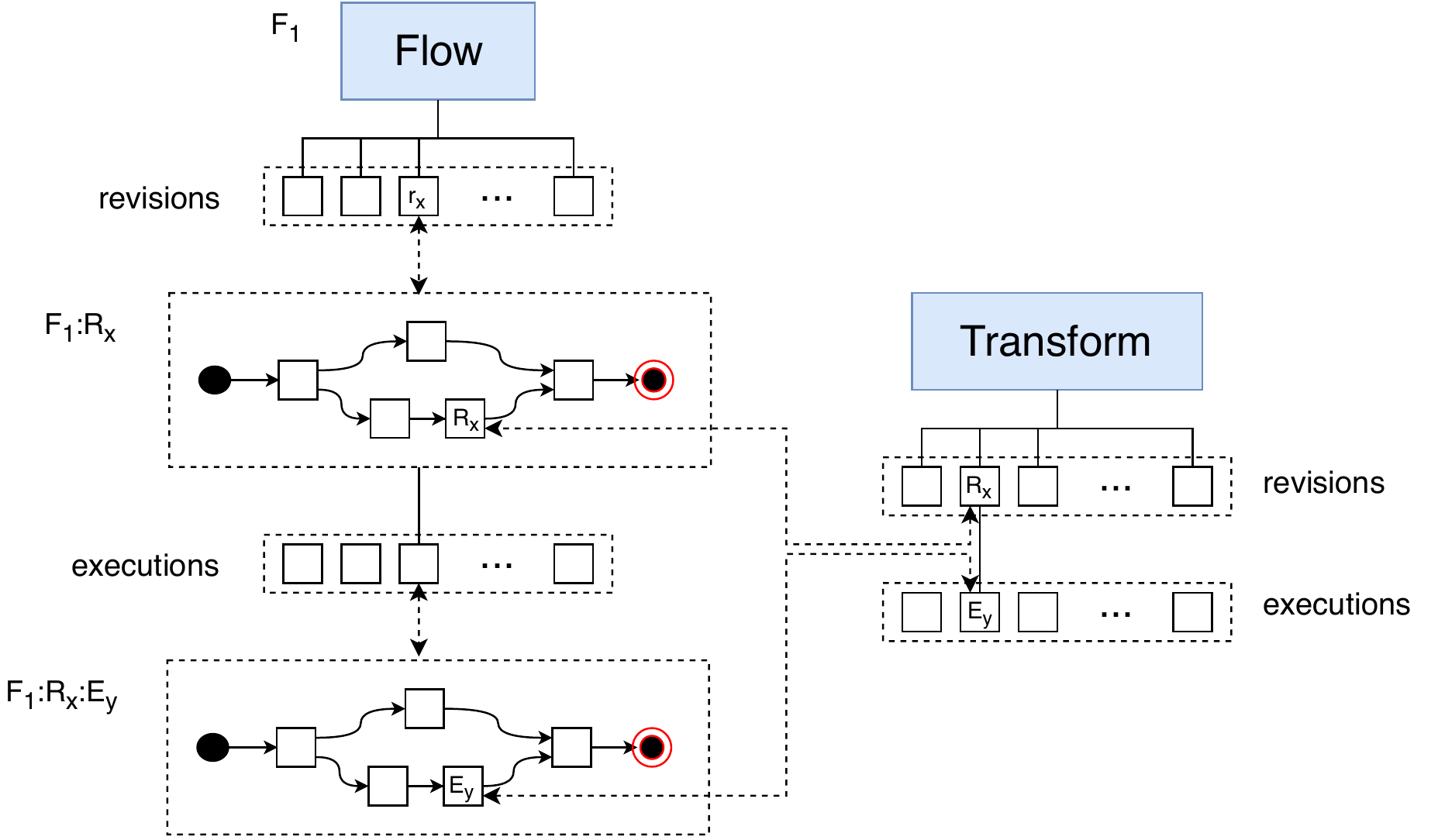}
  \caption{Mechanics of flow execution as it relates to the execution of transformations it consists of.}
  \label{fig:tr_flows}
\end{figure}

As pictured in Figure \ref{fig:tr_flows}, flow revision contains complete definition of a flow (as well as reference to external system formally managing flow instance) referring to transform revisions tied together in a directed graph. In this proposal we omit the details of flow definition DSL (for examples see AWS Step Functions \cite{rel_aws_step_func}, Airflow \cite{rel_airflow}) and focus on abstractions essential for tracking purposes. 

Flow execution records the sequence and details of operations performed when a particular revision of the flow is being run. Flow execution contains the log of all transform executions invoked as a part of it. The format of the log is natural extension of DSL used to define flow itself. 

\subsection{Grouping}

As a tracked system grows in size it becomes important to be able to observe groups of entities as those relate to each other while performing similar functions or as those belong to the same subsystem. The feature of grouping is introduced as a layer on top of the entire model allowing any set of entities to be combined together. Entities (and groups itself) can participate in a number of such formations allowing large models to be broken down into much smaller, better observable and trackable units.  

At a lower level applications can make use of grouping for managing data primitives broken into a collection of otherwise independent datasets. Use cases like that arise when datasets (or other entities) are created on-the-fly with subsequent need for being addressed using a certain query semantics (e.g. sliding window). Grouping those together provides formal mechanism for running such queries. 

Hierarchical grouping capabilities have important implications for data cataloging. Using grouping semantics in this connotation enables designers to combine high- and low-level information about the system in a single consistent location. Properly designed dataset grouping paradigm can enable efficient data discovery, spread awareness, and boost adoption. 

\subsection{Lineage Progression}

Data tracking model does not exist in a static state. It constantly evolves by transforms (and flows) being run, new datasets being added, types registered, etc. In this section we propose scalable mechanism for keeping track of such changes. 

\subsubsection{Transactions}

As defined in this proposal, data lineage tracking system is responsible for recording of changes occurring in live application. Most of such mutations either occur simultaneously or have very strong cause-effect relationship which justifies those being added to the model all at once. In order to facilitate this we introduce the concept of \textbf{transaction} which groups together a set of model modifications and applies those to the model. 

This proposal assumes that all changes to the model are performed in the context of implicit of explicit transaction. Transactions constitute of independent lists of entities being added, modified and removed along with some metadata added for auditability. Proposed mechanism allows applications to define the level of granularity appropriate for a specific us case (transaction size can vary from one item to thousands based on the scale of the system and required detailization level).

\begin{figure}[htbp]
  \centering
  \includegraphics[scale=0.6]{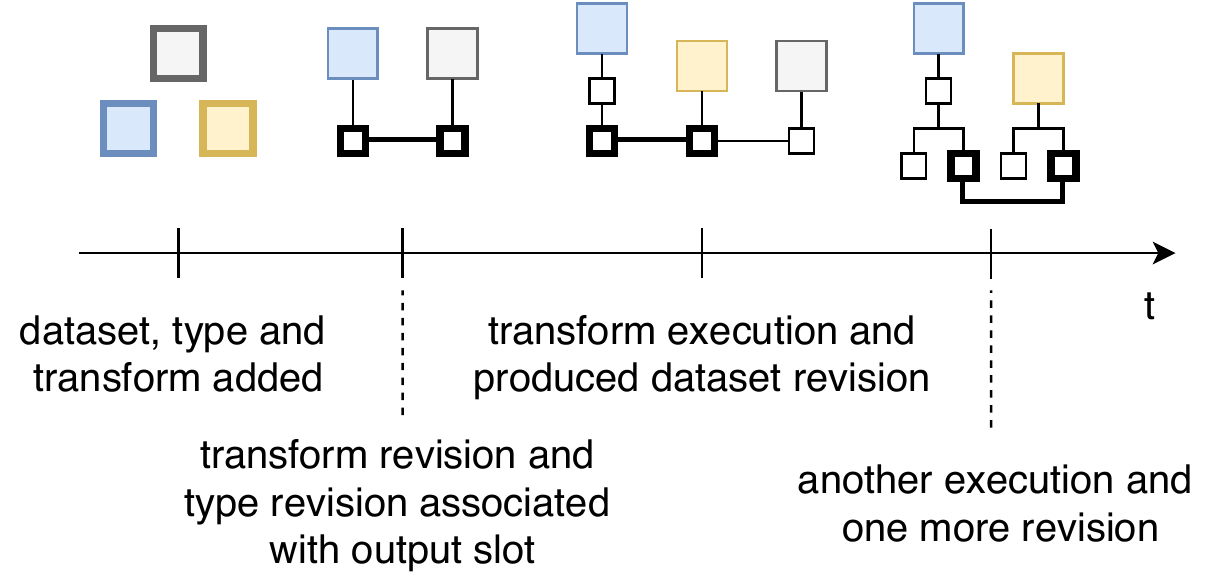}
  \caption{Timeline visualization of transaction log.}
  \label{fig:tr_trans}
\end{figure}

Transaction semantics enables an additional level of auditability by introducing a concept of a changelog consisting of all modifications made to the system. An example shown in Figure \ref{fig:tr_trans} renders side by side the list of transactions made (underneath the timeline) and point in time snapshots of the system at the moments immediately following those commits with transaction effects highlighted (above the timeline). 

\subsubsection{Identity}

The concept of transaction log provides proper level of abstraction for introducing yet another essential piece of data lineage model - identity management. The concept of \textbf{identity} is introduced at transaction level and allows applications to record runtime access details along with each committed transaction. Identity information as it is recorded as a part of transaction log provides important insight for the purposes of security and compliance auditability. 

Identity object associated with each transaction can be adopted for aggregating all of the following properties of surrounding runtime environment. Host applications can choose to separate and record committer identity independently from executor identity. Additionally, properties of respective software systems can be recorded in addition to the identity of authenticated agents (software versions, runtime information, etc.). 

Current proposal resorts to the ability to record the identity of committer along with each transaction and does not follow the thread of verification process that needs to be put in place to ensure correctness of metadata identity entity is populated with. Applications can choose to adopt stronger enforcement paradigm and integrate one of federated authentication mechanisms to eliminate the possibility of maliciously formed transactions to be committed. 

\section{Scaling Considerations}
\label{sec:scaling}

Major goal of the proposed system is to allow organizations or any size to democratize the use of data while preserving security and privacy considerations and compliance conformance as first class citizen concepts of data infrastructure design. 

\begin{figure}[htbp]
  \centering
  \includegraphics[scale=0.6]{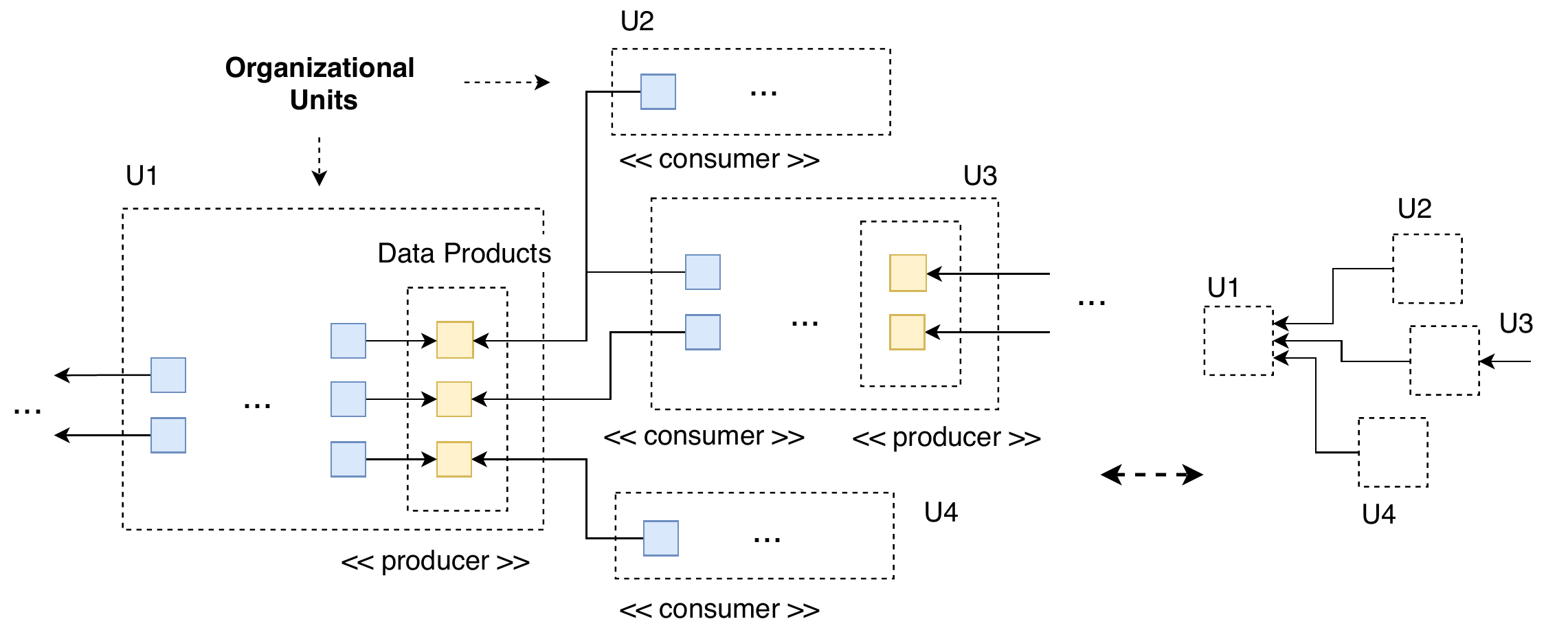}
  \caption{Data producer-consumer model as it scales along with organization size. Detailed view on the left and higher-level picture on the right.}
  \label{fig:scaling_prod_cons}
\end{figure}

Formalizing data flow in complex environment of a large (or even medium-sized) organization proves to be a challenging task. The use of higher-level abstractions in the form of groups or organizational units can introduce proper structure needed to formalize relations between departments of real-world organization. In this paradigm actors participate in data exchange as producers or consumers of data products. The use of such instruments in combination paints bigger picture of data ecosystem established within organization (see Figure \ref{fig:scaling_prod_cons}).

To make the model complete, additional messaging is introduced between actors allowing them to efficiently exchange with updates regarding data feeds available within organization. One of the key concepts of this message exchange becomes the notion of data product and transform deprecation which acts as a mechanism for the organization to move forward as data becomes obsolete, needs to be cleaned up or removed for compliance reasons. 

\begin{figure}[htbp]
  \centering
  \includegraphics[scale=0.6]{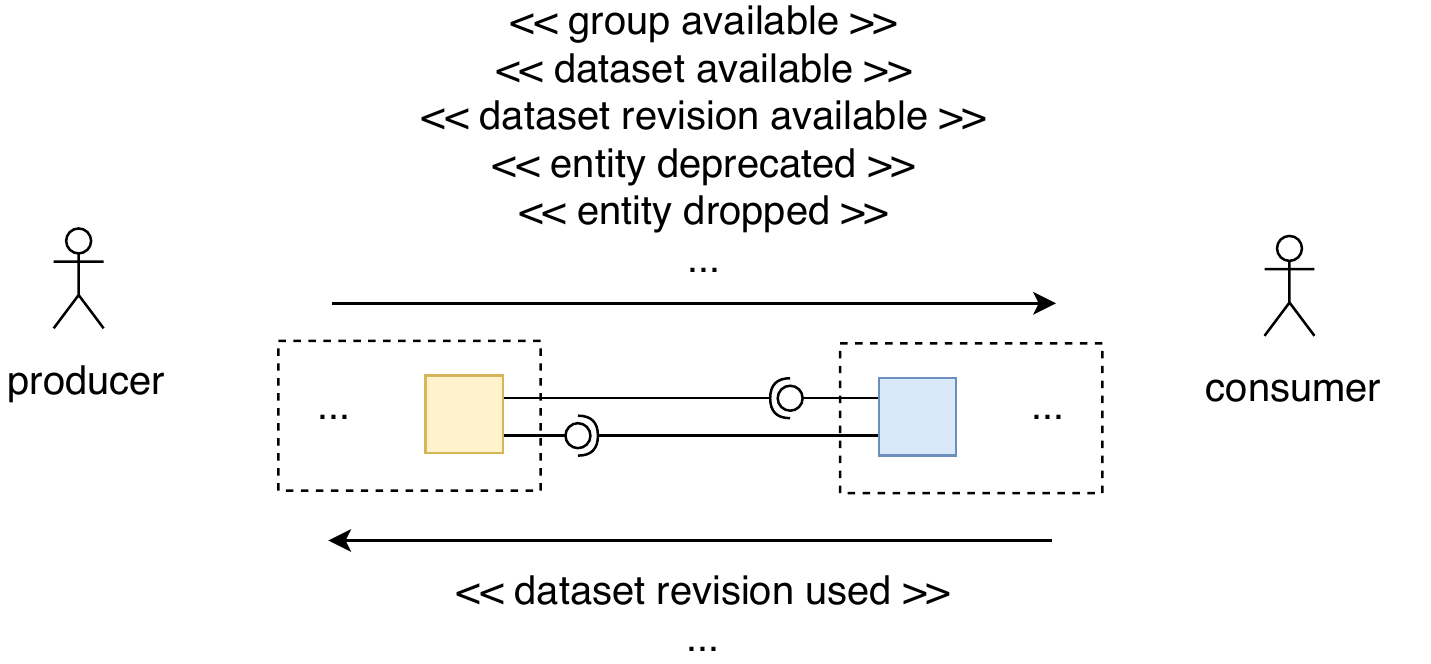}
  \caption{Messaging system for data broadcasting.}
\end{figure}

Host applications can extend the protocol and entity definitions with domain-specific lexicon to implement custom business logic and support internal processes (e.g. use tagged datasets within entity groups to enable filtering and data feed subscriptions within the organization).

\section{Analysis Techniques}
\label{sec:examples}

Proposed theoretical model covers a wide range of workflow types and data processing paradigms. We use this section to illustrate the level of flexibility it offers by covering some specific examples in a greater level of detail. 

\subsection{Tracing Fundamentals}
\label{sec:track_fund}

Ability to establish relationship of data to its ancestors and descendants (we refer to it as data lineage) is a key to a great number of tasks commonly performed in data-driven environments. We separately cover mechanics of forward- and backward-tracing and elaborate on use cases those basic capabilities enable. 

\subsubsection{Forward Tracking}

The need for forward tracing arises when a particular dataset revision or input needs to be traced to all of its consumers. Such operations are common when dealing with sensitive data and compliances (e.g. GDPR). 

\begin{figure}[htbp]
  \centering
  \includegraphics[scale=0.6]{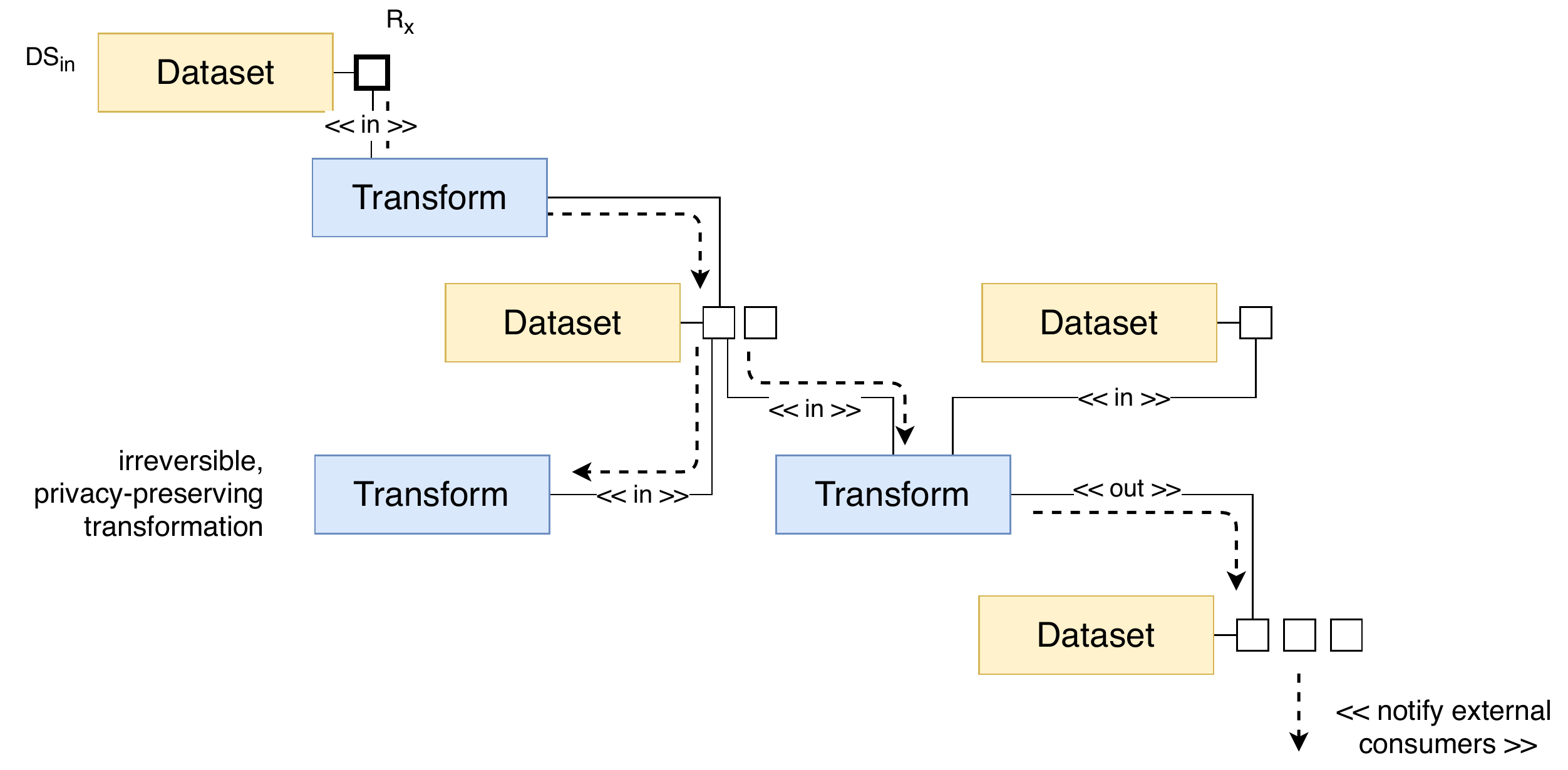}
  \caption{Example of tracing dataset revision propagation through the model.}
  \label{fig:fwrd_tracing}
\end{figure}

To illustrate forward-tracing in action we consider the scenario where revision \(DS_{in}:R_{x}\) goes through the chain of transformations and as a result leaves the trace consisting of a number of derivative data products (see Figure \ref{fig:fwrd_tracing}). Due to the fact that every transform execution contains explicit references to input and output revisions, traversing the graph of those happens to be a trivial task. 

The result of such traversal provides the list of dataset revisions being direct ancestors of \(DS_{in}:R_{x}\) produced by all transforms dependent on it. Note that as involved transforms are being executed over and over again, each respective level of virtual dependency tree grows wider. In the absence of formal tracking mechanism (like the one we propose) this usually leads to numerous branches of this tree being overlooked and ignored which results into violation of at times strict rules and costly consequences for companies operating those data systems.  

The tree of data dependencies can often grow huge in case of successful and very commonly used data products. Downstream application can choose to rely on transform properties indicating whether traversal should continue through it (with the aim of narrowing down the scope of the traversal) or to perform unconditional search throughout the entire hierarchy. Marking transforms as irreversible and privacy-preserving can reduce complexity of search inquiries. At the same time the correctness of these properties is not strongly enforceable and easily provable. 

The resulting set of all data products discovered during forward-tracking can be used for internal data cleanup, adoption analysis, compliance conformance and other important tasks. Leafs of data lineage tree constructed during this operation can be used to notify external consumers of the data in the scenarios where investigation needs to proceed beyond the boundaries of isolated tracking environment.

\subsubsection{Backward Tracking}

On the other hand the need to track specific dataset to its origins arises equally often. This ability is a fundamental building block of experiment reproducibility which proves to be essential in the field of machine learning and a number of domains where frequent experimentation is an integral part of the process (medicine, biology, physics, etc.). 

The other important application of backward tracking is the analysis of ascendants of a particular data revision in general or direct ascendants belonging to a specific dataset in particular. This use case powers dependency management, data changes effect analysis (more on that in Section \ref{sec:tracking_effect}). 

\begin{figure}[htbp]
  \centering
  \includegraphics[scale=0.6]{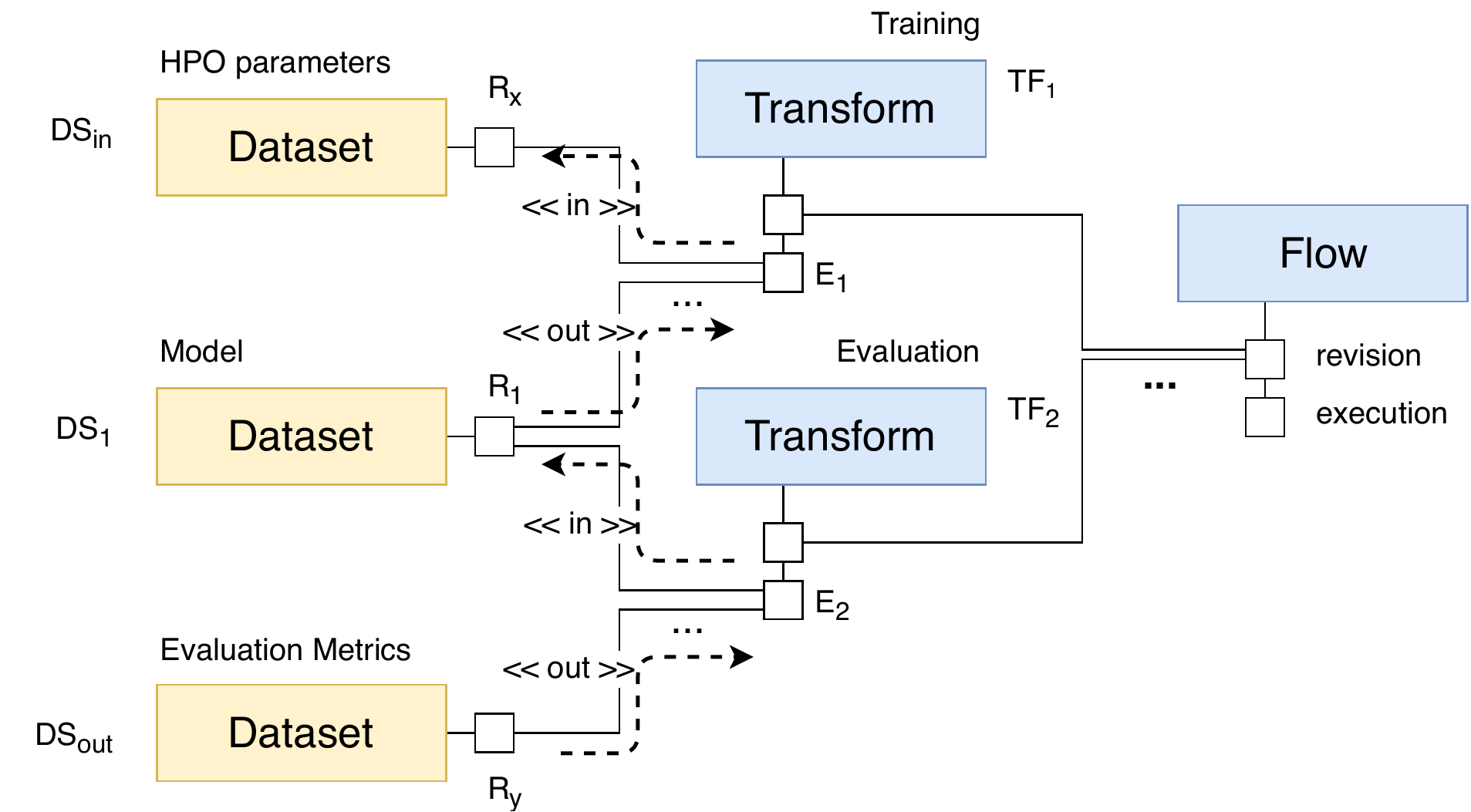}
  \caption{Trivial model training flow used to demonstrate backtracking from evaluation results to HPO parameters.}
  \label{fig:bcwd_tracking}
\end{figure}

For a quick illustration, let's consider a simple example of two-stage machine learning flow consisting of training (\(TF_1\)) and evaluation (\(TF_2\)) transforms (see Figure \ref{fig:bcwd_tracking}). In this example initial dataset \(DS_{in}\) containing training parameters is consumed by \(TF_1\) which produces the model persisted in \(DS_1\). Subsequently model evaluation performed by \(TR_2\) produces metrics stored in \(DS_{out}\). 

In case if specific set of metrics produced by \(TF_2:E_2\) needs to be analyzed, lineage of dataset revision \(DS_{out}:R_y\) can be traced back to a specific ancestor from dataset \(DS_{in}\): 

\begin{equation}
Ancestors(DS_{out}:R_{y}, DS_{in}) = L(DS_{in}:R_x)
\end{equation}

The only route can be found between those two revisions of interest: 

\begin{equation}
LineageRoute(DS_{out}:R_{y}, DS_{in}:R_{x}) = L([TF_1:E_1, DS_1:R_1, TF_2:E_2])
\end{equation}

It is important to underline that due to many to many semantics supported by transformations (as those are defined in this proposal) both \(Ancestors()\) and \(LineageRoute()\) functions can provide zero to many items in response. 

\subsection{Tracking Effect of Changes}
\label{sec:tracking_effect}

Most data lineage analysis use cases are based on the foundation of tracing principles discussed in the previous sections. Here we provide a glimpse into some specific scenarios and show how application-specific logic can be layered on top of core lineage functionality. 

\subsubsection{Impact of Code Changes on Produced Data}

In the most trivial meaning data lineage model can be used for the analysis of causal relationships between code and resulting data changes. The need for such type of analysis appears during evaluation, debugging, and troubleshooting of software components. Ability to efficiently pinpoint a change, resulting in the appearance of observed data phenomena, in the code proves to be of significant value to system maintainers. 

\begin{figure}[htbp]
  \centering
  \includegraphics[scale=0.6]{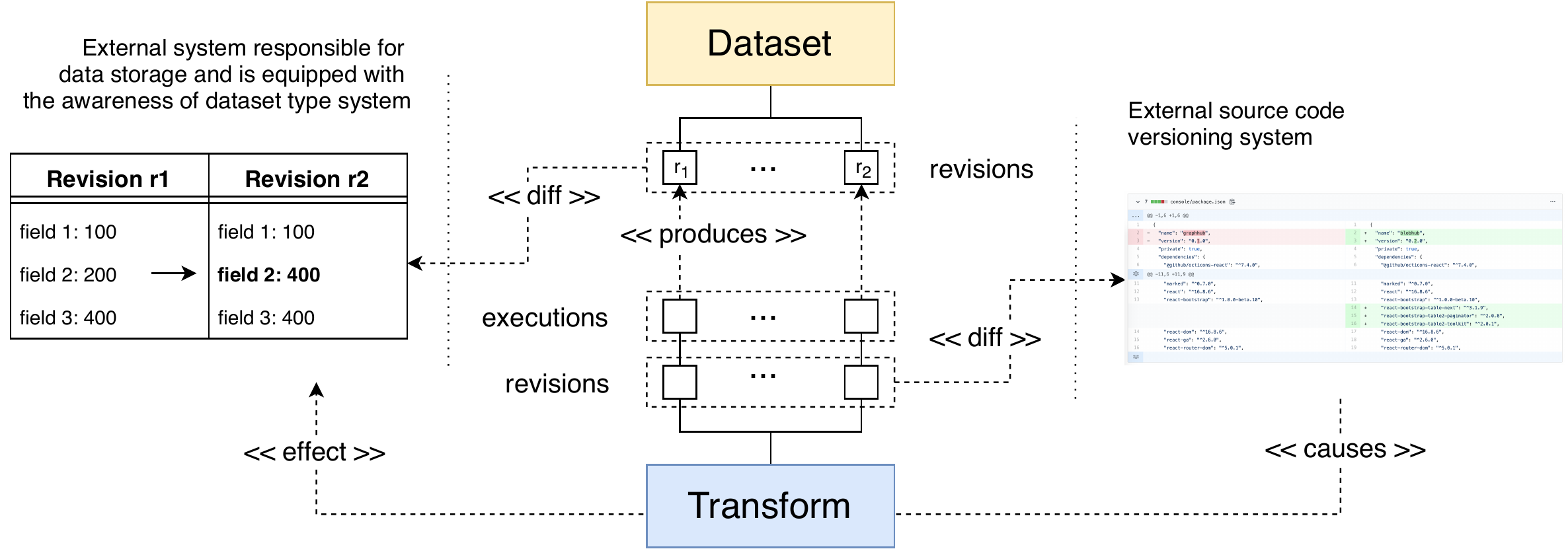}
  \caption{Illustration of the impact a specific code change (on the right) makes on the result in the metric of interest (on the left).}
  \label{fig:code_to_data}
\end{figure}

An important aspect of current proposal (that has been covered in Sections \ref{sec:datasets} and \ref{sec:transforms}) is the interface between lineage system and parent application. As explained in Figure \ref{fig:code_to_data}, application leverages external tools for analysis and visualization of code, data state, and changes. Lineage system is used for reference tracking - tying things together and maintaining relationships. This separation of responsibilities proves itself especially powerful when complex multi-tiered applications relying on a variety of data storage platforms and execution engines. Those heterogeneous environments do not naturally come with mechanisms for in-depth analysis which complicates its support and evolution. Lack of visibility into the behavioral aspects of such systems results in costly mistakes that are not easy to correct. The approach discussed in this proposal is focused on introducing preventive techniques focused on elimination of ambiguity, black boxes and vulnerabilities in complex data processing environments. 

In the example presented here transform and dataset are linked directly (transform executions produce revisions of the dataset). In practice those can be separated by multiple layers preserving the same ability to track changes as those propagate through the system.  

\subsubsection{Impact of Data Changes on Produced Data}

As an extension of the example reviewed above we evaluate how trend analysis can be performed with the help of tracing capabilities provided by the model. We consider a scenario where a set of model training parameters affects evaluation metrics of the model trained by a multi-step workflow (see Figure \ref{fig:code_to_data}). 

A number of important concepts are illustrated in this example. Previously considered tracing algorithm is used for establishing relationships between dataset revisions of \(DS_{in}\) and \(DS_{out}\). Once those are established, application interprets data associated with \(DS_{out}\) revisions for analysis (which in this case in the visualization) of the effect changes in HPO parameters make. 

\begin{figure}[htbp]
  \centering
  \includegraphics[scale=0.6]{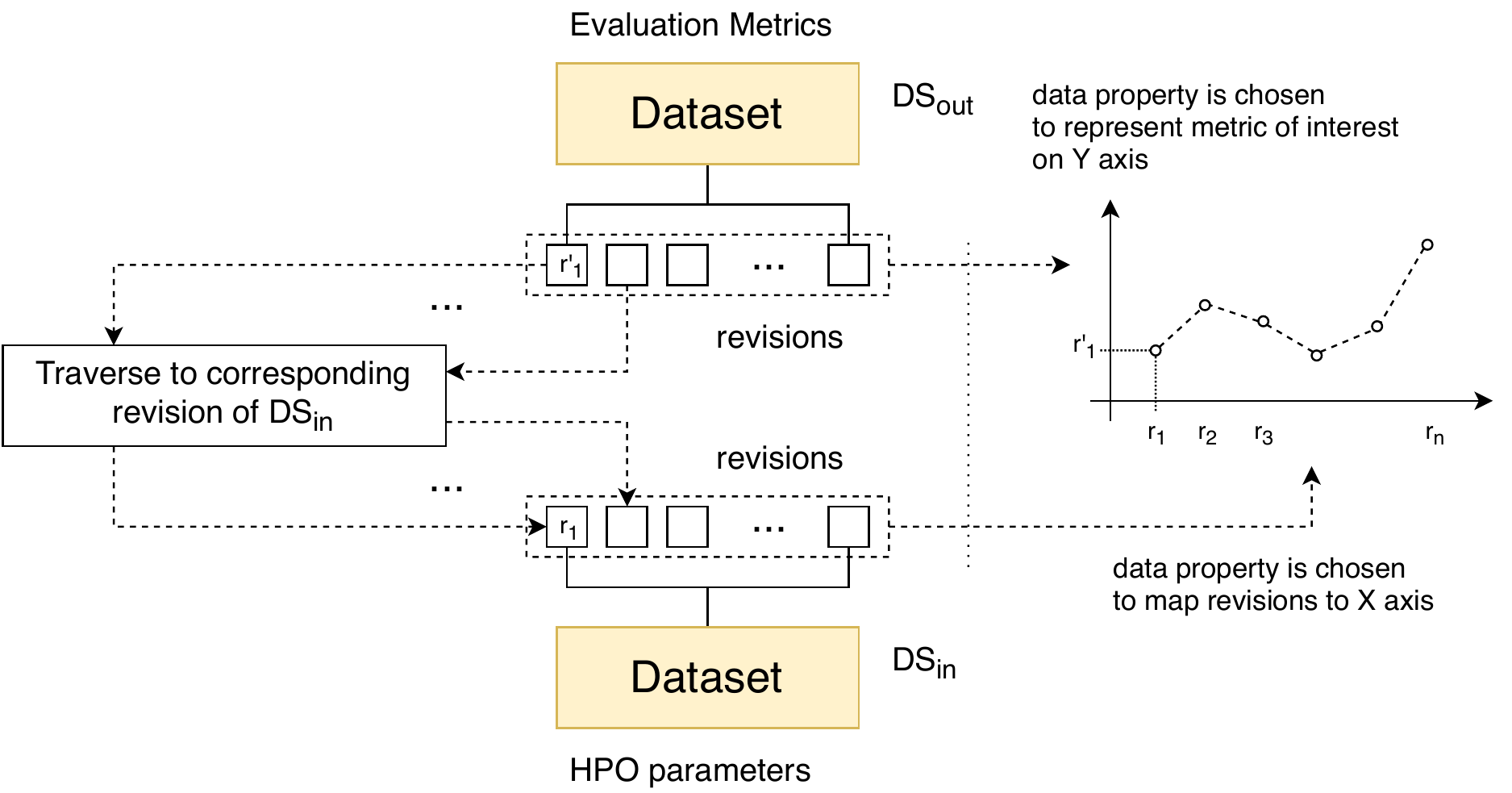}
  \caption{Visualization of the effect changes in one dataset (DS\textsubscript{in}) have on the other one (DS\textsubscript{out}).}
  \label{ref:data_to_data}
\end{figure}

In here we emphasise on the fact that external systems can leverage from its data format and meaning awareness capabilities and tooling (visualization, alerting) for creating complex workflows powered by causal relationships extracted from lineage layer. 

\subsection{Benign Machine Learning Life Cycle}

Recent rise of machine learning techniques (in areas of vision and natural language processing) resulted in the evolution and appearance of the whole range of data processing workflows (data programming \cite{proc_data_progr}, code-as-supervision \cite{proc_code_as_superv}, federated learning \cite{proc_fed_learning}). The efficiency of adoption of machine learning tools, being a data hungry processes by nature, is defined through the ability to organize and properly manage data used for training and evaluating machine learning models.

\begin{figure}[htbp]
  \centering
  \includegraphics[scale=0.6]{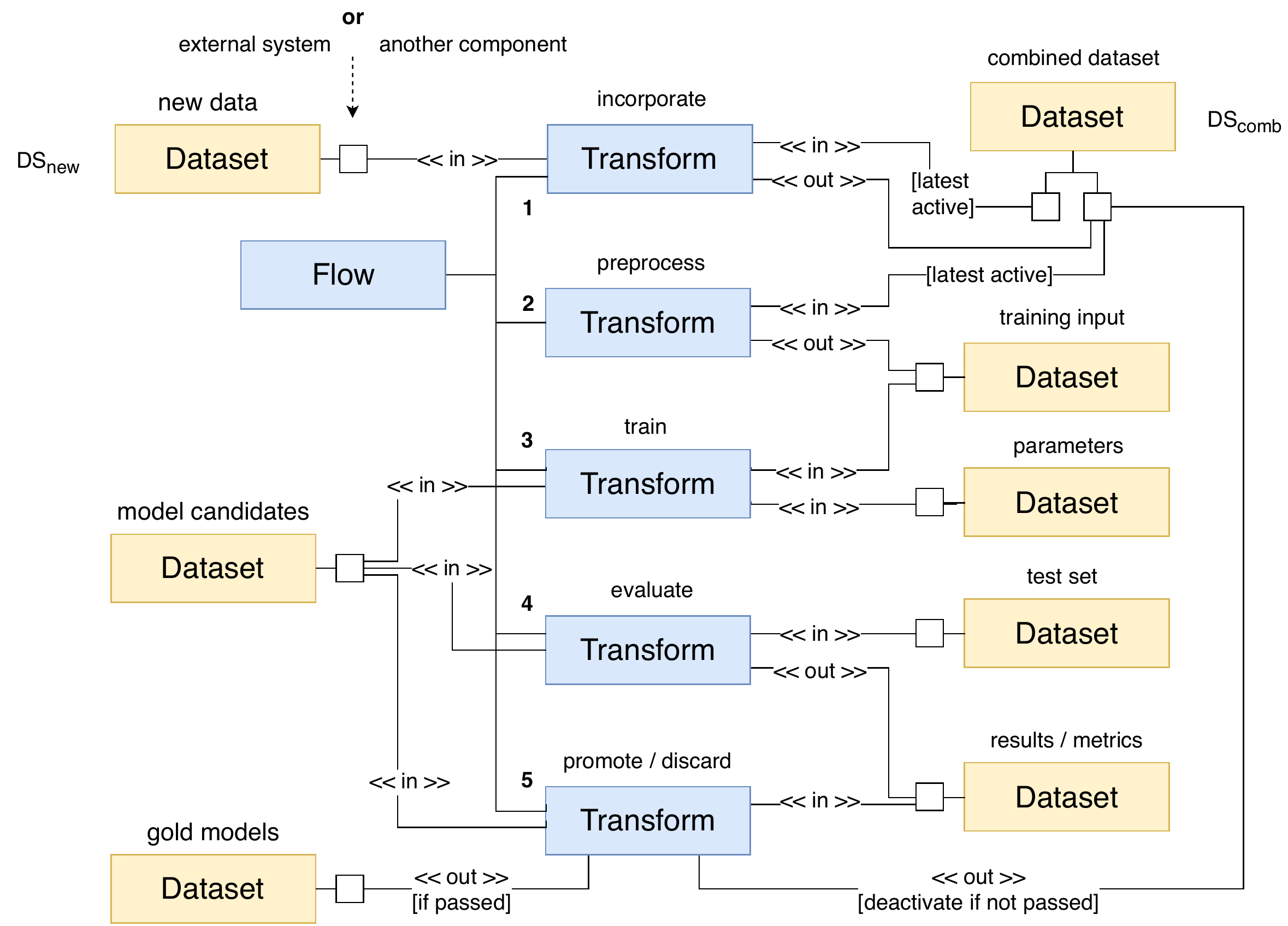}
  \caption{Example machine learning model training flow.}
  \label{fig:ml_flow}
\end{figure}

For the purpose of the example we evaluate simple sequential flow consisting of a set of steps involved into production of models that are considered for production deployment. The flow illustrated in Figure \ref{fig:ml_flow} demonstrates how continuous stream of newly labeled data (tracked within \(DS_{new}\)) is being merged into combined dataset (\(DS_{comb}\) which is used for training and further validation of a model. Newly trained model is being promoted to deployment candidate upon successful validation. Alternatively newly added dump of data is being discarded. 

Ability to freely experiment and iterate on model, parameters and data knowing that it is always possible to establish dependencies between any data products and involved transforms can greatly impact pace and efficiency of machine learning research. In this example, every piece of data being operated by this simple flow is being tracked in the form of dataset revision, which in conjunction with recording invocations of all pieces of logic as transform executions allows following its lineage all the way through in both directions of inputs used and outputs produced. 

In addition to traceability itself, proposed paradigm allows reconstructing complete lineage of data and transforms involved in the production of a specific modeling artifact. The development of any machine learning product critically depends on the ability to reproduce and iterate on the results of a particular experiment (granted that a lot of factors mentioned previously are contributing to some of the key components of it being non-deterministic by nature). 

\subsection{Data Modality Transition}
\label{sec:data_mod_tr}

Applications operate with data materialized in a number of forms including: streams, relational tables, key-value spaces, blob repositories. Real life systems usually involve operating on the boundaries of those modalities constantly shifting data across those to perform specific business functions. 

\begin{figure}[htbp]
  \centering
  \includegraphics[scale=0.6]{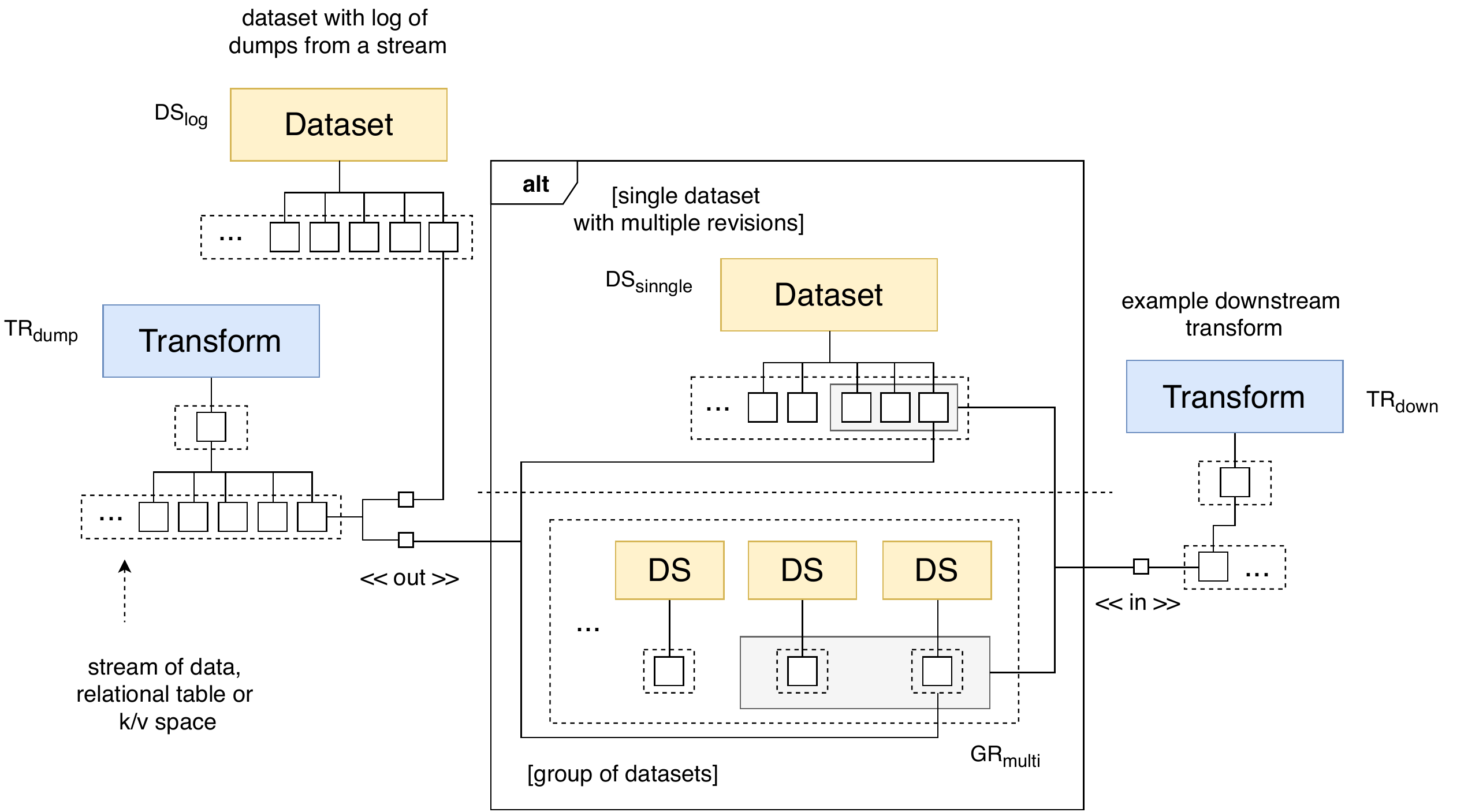}
  \caption{Stream/table data processing flow.}
  \label{fig:data_mod_tr}
\end{figure}

The fluidity of those processes requires separate considerations and is not fully explored as a part of this work. However proposed paradigm allows capturing enough metadata for those processes to be properly controlled and audited. 

Consider the scenario where a system produces periodic dumps of data from a stream into blob store that is being further consumed (with some sliding window) for downstream analysis (see Figure \ref{fig:data_mod_tr} for details). This trivial use case involves the collection of some basic metadata associated with each dump (\(DS_{log}\) is introduced for this purpose) and dumps itself. Data model designers have flexibility of representing dumps with either single dataset containing multiple revisions (\(DS_{single}\)) or multiple single-revisioned datasets grouped together (\(GR_{multi}\)). Each execution of downstream transform (\(TR_{down}\)) refers to a specific subset of those revisions for the analysis it performs. Exact list (or range) of consumed revisions is associated with corresponding execution of \(TR_{down}\).

As a result, a combination of \(TR_{dump}\), \(DS_{log}\), \(DS_{single}\) (or \(GR_{multi}\)), and \(TR_{down}\) provides comprehensive view into the flow of every bit of data as it crosses the boundaries of the original stream and proceeds to downstream components. 

\section{Conclusion}
\label{sec:conclusion}

In this paper we defined the problem of reconstructing data lineage and briefly evaluated state of the ecosystem of tools and services (open source and proprietary) available around it. We proposed a design of a system capable of addressing the most topical challenges associated with rapid growth of data and ways it is used in infrastructures of any size. 

The framework proposed in this work provides theoretical basis for representing existing (and designing new) data flows in a formal manner suitable for collection and analysis. The toolbox we defined is based on the foundation of datasets and transforms extended with ability to record slot affiliation, type information, and iterative nature of existing software systems. 

We acknowledge that with this work we barely scratch the surface of data traceability and lineage reconstruction problem. Further research can be performed in the directions of covering cross-modal data transitions; ensuring correctness of submitted changes; data ownership, access control and active identity verification of involved agents. In its current state proposed methodology is focused on passive data collection. Areas of controlling (or informing) execution of runtime data processing environment and building predictive capabilities on top of lineage system can be further explored.  

One of the driving forces behind this work is the desire to facilitate further research in this direction and encourage standardization of formal data flow tracking approaches. The work presented in this paper can act as a starting point for an open specification of data lineage and tracing system. Collaborative approach to defining common interfaces can further speedup the adoption of lineage methodology throughout the industry.

\printbibliography

\appendix

\section{Programming Paradigm}
\label{sec:progr}

Implementation details of lineage system described in this paper are deemed to be trivial and are not covered in here for brevity. However, to make an overview of proposed approach complete we provide some remarks regarding its public interface and potential integration options.

One of the key concepts of this proposal is the idea that data lineage system itself does not provide neither data storage nor execution environment. The reasoning behind this approach is described in earlier sections. Interfaces of the system are shaped by this decision and focus on recording of data manipulations and performed executions in a passive manner. This leads to two potential integration scenarios that are being explored here. 

\subsection{Explicit Use}

The interface directly exposed by the system provides low-level and granular access to all of its entities. The reliance on this layer provides the highest level of flexibility at the expense of having highly verbose way of expressing occurring changes (see Listing \ref{lst:expl_use}). 

\begin{lstlisting}[
  caption={Manual recording of an execution.},
  style={standalone},
  label={lst:expl_use}
]
# Find dependencies 
tr_rev = model.find_transform(id=...).latest()
ds_rev_in_1 = model.find_dataset(id=...).latest()
ds_rev_in_2 = model.find_dataset(id=...).revision(id=...)
ds_out_1 = model.find_dataset(id=...)
ds_out_2 = model.find_dataset(id=...)

# Record outputs 
ds_rev_out_1 = %\textbf{ds\_out\_1.new\_revision}%(
  external_blob_id=...)
ds_rev_out_2 = %\textbf{ds\_out\_2.new\_revision}%(
  external_blob_id=...)
  
# Record transform execution
tr_exec = %\textbf{tr\_rev.new\_execution}%(input={
  tr_rev.input.slots[0].id: ds_rev_in_1, 
  tr_rev.input.slots[1].id: ds_rev_in_2
}, output={
  tr_rev.output.slots[0].id: ds_rev_out_1, 
  tr_rev.output.slots[1].id: ds_rev_out_2
})

# Commit transaction consisting of new execution and its products
transaction = %\textbf{model.new\_transaction()}%
transaction.%\textbf{add\_transform\_execution}%(tr_exec)
transaction.%\textbf{add\_dataset\_revision}%(ds_rev_out_1)
transaction.%\textbf{add\_dataset\_revision}%(ds_rev_out_2)
%\textbf{transaction.commit}%()
\end{lstlisting}

\begin{figure}[htbp]
  \centering
  \includegraphics[scale=0.6]{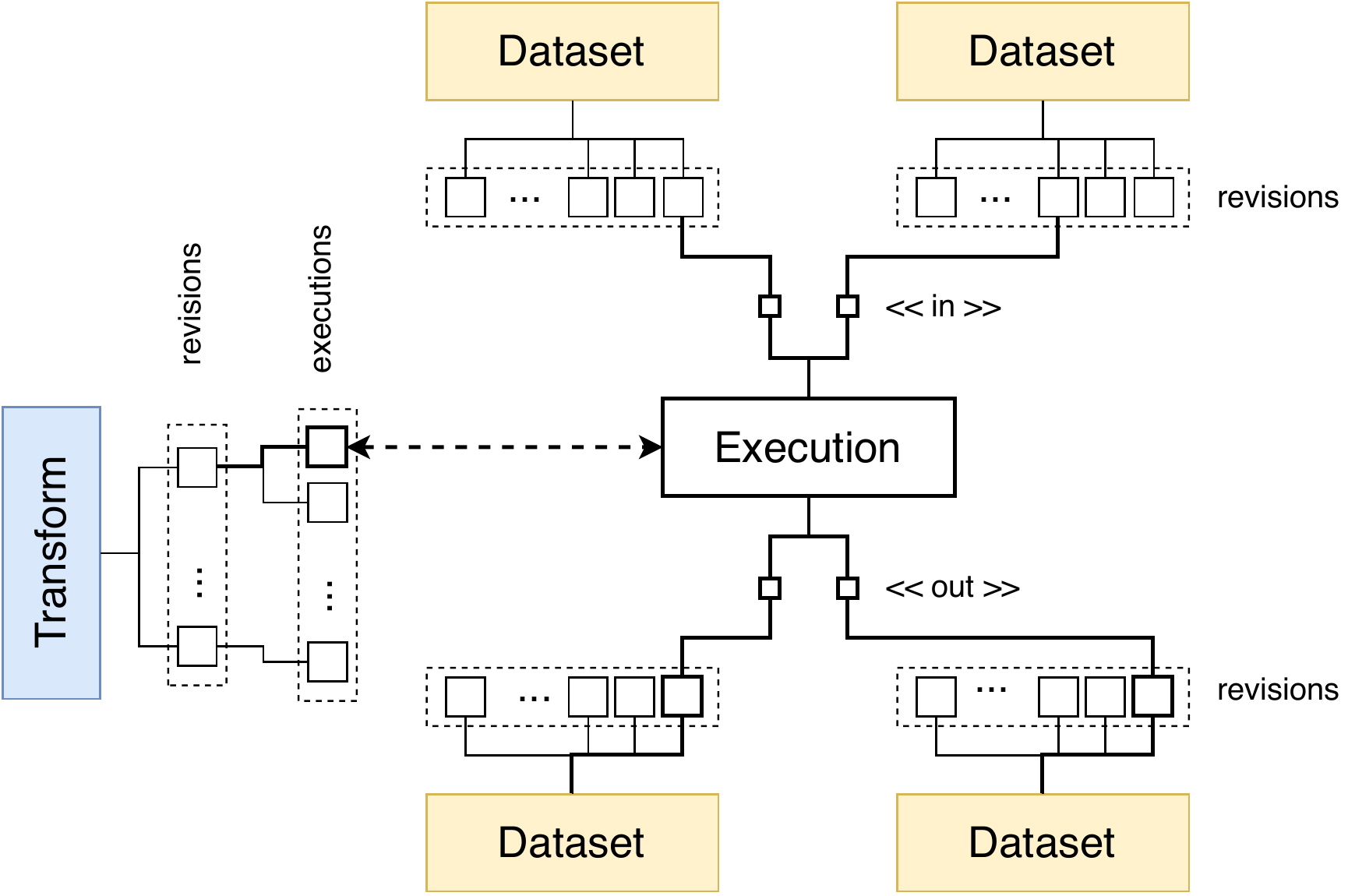}
  \caption{Effect of a transaction constructed manually.}
  \label{fig:expl_effect}
\end{figure}

The snippet makes a number of assumptions regarding input and output structure of the transform being executed for illustration purposes. In practice, calling application will likely populate this information dynamically using embedded self-awareness mechanisms to make sure this reflects its actual state at the moment of recording. The results of snippet execution and the impact it makes on the model are illustrated in Figure \ref{fig:expl_effect}. 

One of the fundamental principles of the proposed paradigm, however, is the fact that data engineers do not have explicitly write the code above to record data transformations. Deep integration of the system into APIs used to perform transformations itself is essential for usability and reliability aspects of this proposal. Ad hoc tracking of all performed operations is not deemed to be neither efficient nor desirable from a consumer point of view.

\subsection{Implicit Use}

\begin{figure}[htbp]
  \centering
  \includegraphics[scale=0.6]{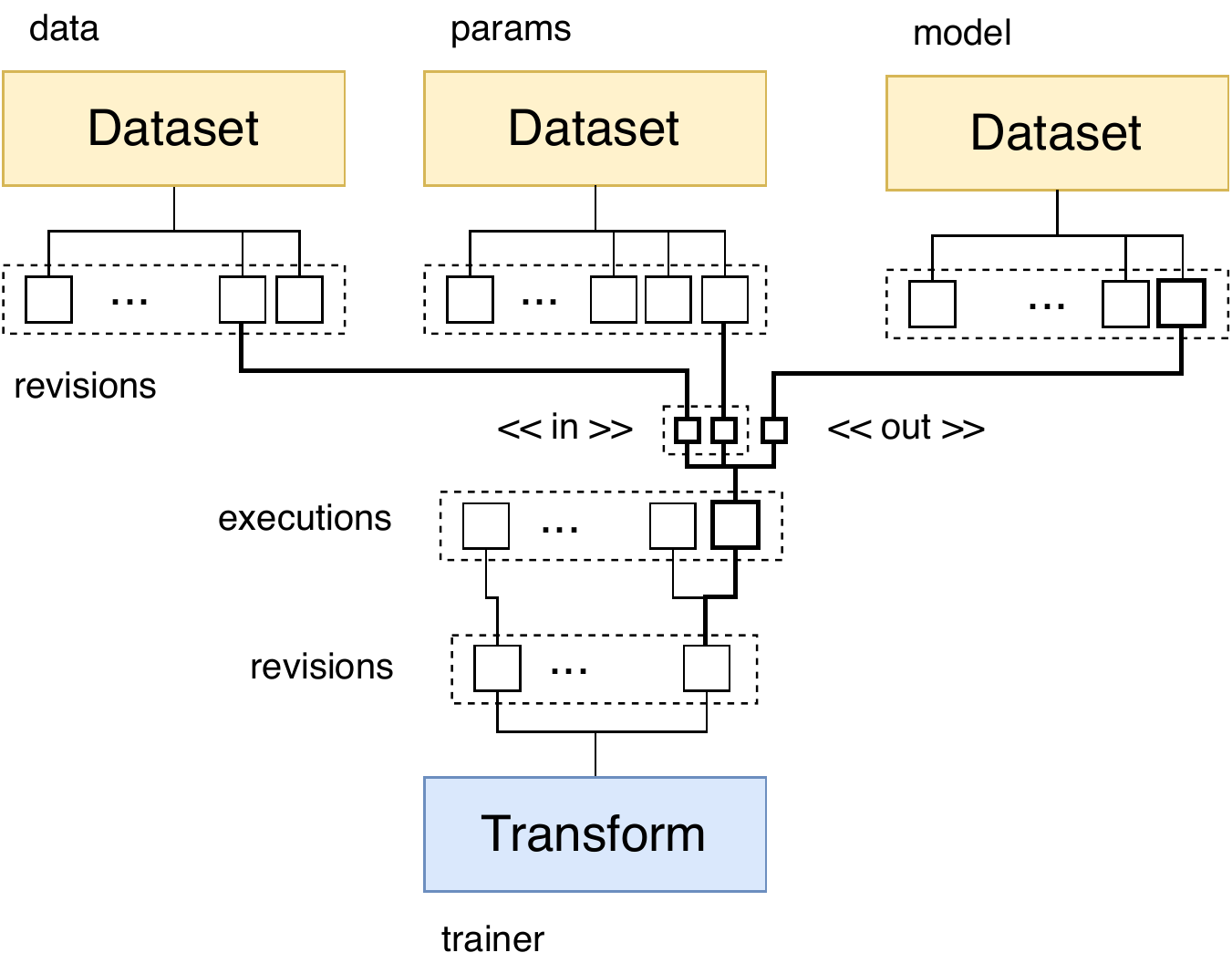}
  \caption{Effect of implicitly populated transaction.}
  \label{fig:impl_effect}
\end{figure}

Building on the foundation of low-level API described above, we demonstrate the impact of deep integration into data storage, processing, and DNN frameworks. Table \ref{table:impl_use} builds parallel between the code naturally produced during experimentation and tracking logic invoked under the covers. Figure \ref{fig:impl_effect} confirms that the same level of detail is achievable even if tracking subsystem is completely abstracted.

\lstset{style=inplace}
\begin{table}
  \centering
  \begin{tabular}{p{0.3\linewidth}p{0.6\linewidth}}
\toprule  
Written code & Internal logic \\
\midrule
\begin{lstlisting}
data = Data.load(...)
\end{lstlisting} & 
\begin{lstlisting}
ds_rev_in_1 = model
  .find_dataset(id=...)
  .revison(id=...)
\end{lstlisting} \\

\begin{lstlisting}
params = Data.load(...)
\end{lstlisting} & 
\begin{lstlisting}
ds_rev_in_2 = model
  .find_dataset(id=...)
  .revison(id=...)
\end{lstlisting} \\

\begin{lstlisting}
model = Model()
\end{lstlisting} & 
\begin{lstlisting}
ds_out_1 = model.find_dataset(id=...)
ds_rev_out_1 = %\textbf{ds\_out\_1.new\_revision()}%
\end{lstlisting} \\    

\begin{lstlisting}
trainer = Trainer()
\end{lstlisting} & 
\begin{lstlisting}
tr_rev = model.find_transform(id=...).latest()
\end{lstlisting} \\    

\begin{lstlisting}
trainer.fit(
  model,
  data, 
  params)
\end{lstlisting} & 
\begin{lstlisting}
tr_exec = %\textbf{tr\_rev.new\_execution}%(input={
  tr_rev.input.slots[0].id: ds_rev_in_1, 
  tr_rev.input.slots[1].id: ds_rev_in_2
}, output={
  tr_rev.output.slots[0].id: ds_rev_out_1
})
\end{lstlisting} \\   

\begin{lstlisting}
model.save(...)
\end{lstlisting} & 
\begin{lstlisting}
%\textbf{ds\_rev\_out\_1.set\_external\_blob}%(
  external_blob_id=...)
\end{lstlisting} \\ 

& 
\begin{lstlisting}
transaction = %\textbf{model.new\_transaction}%()
transaction.%\textbf{add\_transform\_execution}%(tr_exec)
transaction.%\textbf{add\_dataset\_revision}%(ds_rev_out_1)
%\textbf{transaction.commit()}%
\end{lstlisting} \\  

\bottomrule
  \end{tabular}
  \caption{Deep integration as demonstrated using machine learning model training example: the code written by the experimenter (on the left); related snippets of data lineage interface integrated into underlying libraries (on the right).}
  \label{table:impl_use}
\end{table}

This integration method resembles the paradigm of configurable logging currently supported by the majority of software frameworks. In this case the interface of data lineage system is being implemented externally and is propagated by application logic to all of its internal components and underlying frameworks for recording purposes. This approach allows combining seamless experimentation with desired level of traceability and reproducibility. 

\section{Data Model Reference}
\label{sec:reference}

\begin{figure}[htbp]
  \centering
  \includegraphics[scale=0.6]{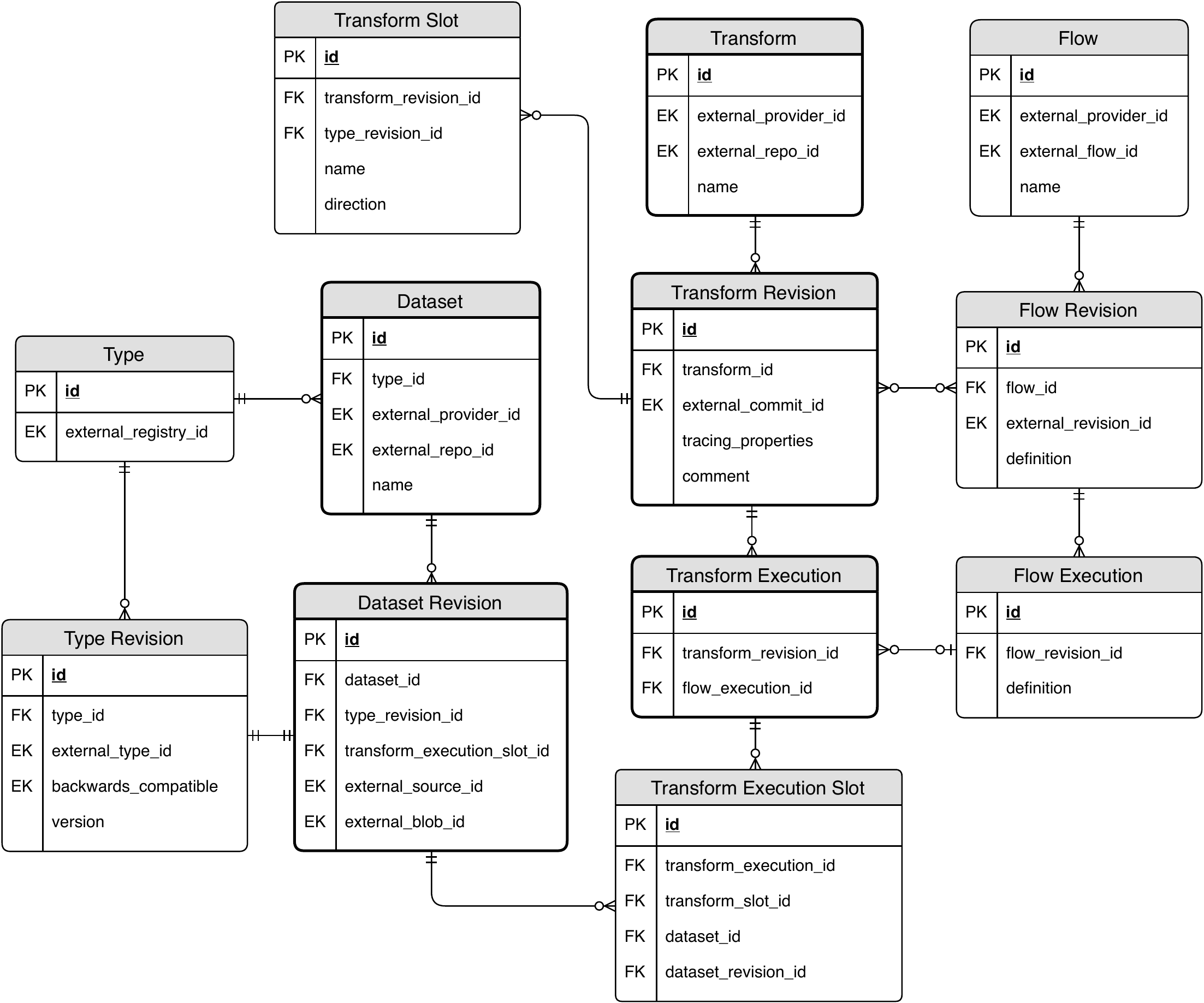}
  \caption{Data Model ERD. Only most significant relations are depicted for clarity.}
  \label{fig:ref_all}
\end{figure}

\begin{figure}[htbp]
  \centering
  \includegraphics[scale=0.6]{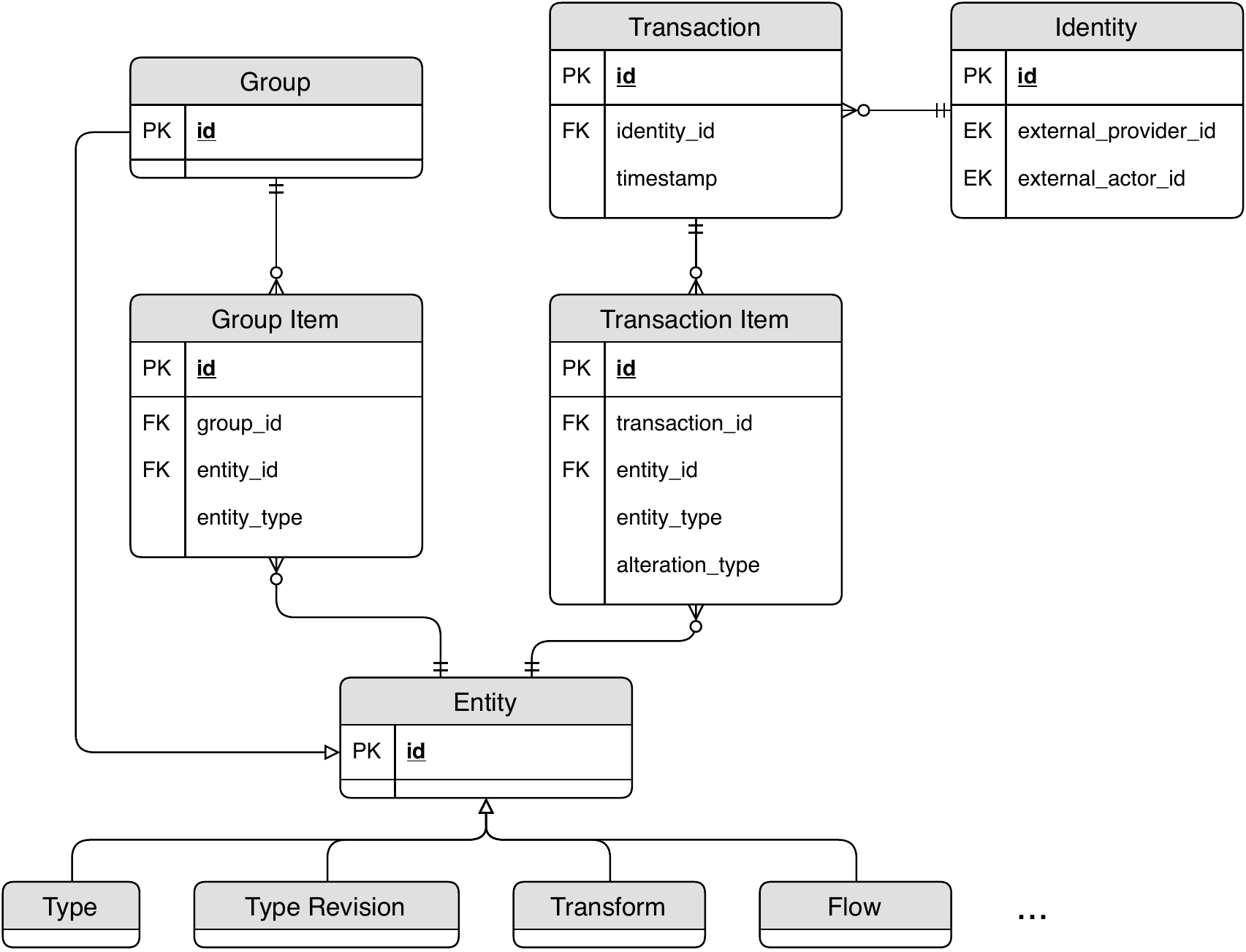}
  \caption{Grouping semantics.}
  \label{fig:ref_groups}
\end{figure}

This section provides a summary of all entities introduced in the paper. Proposed reference acts as a basic skeleton, which should be extended as needed by specific implementations of the model. Most basic entities and are represented in Figure \ref{fig:ref_all}. Additionally we introduce two paradigms used for aggregation of entities and changes being tracked by the model: groups and transactions (see Figure \ref{fig:ref_groups}). 

The remainder of the paper (Subsections \ref{sec:ref_entity} - \ref{sec:ref_transaction}) is devoted to detailed reference that can be used as a starting point for formal data lineage specification. 

\subsection{Entity}
\label{sec:ref_entity}

Entity is a base class for all other primitives described further down below. 

\begin{table}[!htbp]
  \centering
  \begin{tabular}{p{0.1\linewidth}p{0.1\linewidth}p{0.6\linewidth}}
    \toprule
    Key & Name & Description \\
    \midrule
    pk & id & Unique identifier. \\
    & deprecated & Indicates that the entity has been deprecated and reliance on it should no longer be extended. Transform maintainers should be investing into ongoing effort of migrating off deprecated dependencies. This property can be optionally accompanied with termination SLA. \\
    \bottomrule
  \end{tabular}
  \caption{Entity reference.}
\end{table}

\subsection{Type (Entity)}

Type object is used to represent data format. Types (like datasets) do not refer to format definitions but rather act as containers of type revisions. 

\begin{table}[!htbp]
  \centering
  \begin{tabular}{p{0.25\linewidth}p{0.2\linewidth}p{0.4\linewidth}}
    \toprule
    Key & Name & Description \\
    \midrule
    ek (External Registry) & external\_registry\_id & Reference to external registry containing type definitions. \\
    & name & Type name. \\
    \bottomrule
  \end{tabular}
  \caption{Type reference.}
\end{table}

\subsection{Type Revision (Entity)}

The history of data format changes is represented in the form of type revisions. 

\begin{table}[!htbp]
  \centering
  \begin{tabular}{p{0.23\linewidth}p{0.22\linewidth}p{0.4\linewidth}}
    \toprule
    Key & Name & Description \\
    \midrule
    fk (Type) & type\_id & Parent type identifier. \\
    ek (External Type) & external\_type\_id & Type revision identifier in external registry. \\
    & backwards\_compatible & Indicates whether revision preserves backwards compatibility of the format. \\
    & version & Human-readable type version (e.g. "1.0", "1.2.3", "2.1"). \\
    \bottomrule
  \end{tabular}
  \caption{Type Revision reference.}
\end{table}

\subsection{Dataset (Entity)}
\label{sec:ref_dataset}

Dataset refers to a single, indivisible piece of data tracked by the system. 

\begin{table}[!htbp]
  \centering
  \begin{tabular}{p{0.25\linewidth}p{0.2\linewidth}p{0.4\linewidth}}
    \toprule
    Key & Name & Description \\
    \midrule
    fk (Type) & type\_id & Data type identifier. \\
    ek (External Provider) & external\_provider\_id & Identifies external provider where dataset is stored (e.g. Amazon S3 \cite{rel_amazon_s3}, Google Cloud Storage \cite{rel_gcp_storage}). \\
    ek (External Repository) & external\_repo\_id & Identifies repository name (e.g. bucket name). \\
    & name & Dataset name. \\
    \bottomrule
  \end{tabular}
  \caption{Dataset reference.}
\end{table}

\subsection{Dataset Revision (Entity)}
\label{sec:ref_dataset_revision}

Dataset revision stores specific revision of the dataset. 

\begin{table}[!htbp]
  \centering
  \begin{tabular}{p{0.28\linewidth}p{0.28\linewidth}p{0.34\linewidth}}
    \toprule
    Key & Name & Description \\
    \midrule
    fk (Dataset) & dataset\_id & Identifies parent dataset. \\
    fk (Type Revision) & type\_revision\_id & Identifies data type revision. \\
    fk (Transform Execution Slot) & transform\_execution\_slot\_id & Identifier of output slot of transform execution that produced the revision. \\
    ek (External Source) & external\_source\_id & Identifies external source of imported dataset revisions. The field is applicable to revisions registered outside of the scope of a transform execution. \\
    ek (External Blob) & external\_blob\_id & Identifier of dataset stored within external repository (e.g. blob identifier or bucket prefix). \\
    \bottomrule
  \end{tabular}
  \caption{Dataset Revision reference.}
\end{table}

\subsection{Transform (Entity)}
\label{sec:ref_transform}

Represents unit of application logic that can be invoked and versioned independently. 

\begin{table}[!htbp]
  \centering
  \begin{tabular}{p{0.25\linewidth}p{0.2\linewidth}p{0.4\linewidth}}
    \toprule
    Key & Name & Description \\
    \midrule
    ek (External Provider) & external\_provider\_id & Identifies external source version control system (e.g. GitHub \cite{rel_github}, AWS CodeCommit \cite{rel_aws_codecommit}). \\
    ek (External Repository) & external\_repo\_id & Identifies specific code repository. \\
    & name & Transform name. \\
    \bottomrule
  \end{tabular}
  \caption{Transform reference.}
\end{table}

\subsection{Transform Revision (Entity)}
\label{sec:ref_transform_revision}

Transform revisions are used to represent changes occurring in transform logic. Type revisions are logically equivalent to commits in existing source version control systems. 

\begin{table}[!htbp]
  \centering
  \begin{tabular}{p{0.25\linewidth}p{0.2\linewidth}p{0.4\linewidth}}
    \toprule
    Key & Name & Description \\
    \midrule
    fk (Transform) & transform\_id & Identifies parent transform. \\
    ek (External Commit) & external\_commit\_id & Identifies commit in external system. \\
    & tracing\_properties & Bag of properties hinting various aspects of data propagation through the transform. \\
    & comment & Summary of the changes introeuced by the revision. \\
    \bottomrule
  \end{tabular}
  \caption{Transform Revision reference.}
\end{table}

\subsection{Transform Slot (Entity)}
\label{sec:ref_transform_slot}

Transform slots are associated with revisions and record structure and format of input and output data transform revision operates with. Transform slots do not refer to data products directly. 

\begin{table}[!htbp]
  \centering
  \begin{tabular}{p{0.25\linewidth}p{0.25\linewidth}p{0.4\linewidth}}
    \toprule
    Key & Name & Description \\
    \midrule
    fk (Transform Revision) & transform\_revision\_id & Identifier of parent transform revision. \\
    fk (Type Revision) & type\_revision\_id & Revision of the data format the slot operates with. \\
    & name & Slot name. \\
    & direction & Separates input and output slots. \\
    \bottomrule
  \end{tabular}
  \caption{Transform Slot reference.}
\end{table} 

\subsection{Transform Execution (Entity)}
\label{sec:ref_transform_execution}

Describes execution of the transform revision. 

\begin{table}[!htbp]
  \centering
  \begin{tabular}{p{0.25\linewidth}p{0.25\linewidth}p{0.4\linewidth}}
    \toprule
    Key & Name & Description \\
    \midrule
    fk (Transform Revision) & transform\_revision\_id & Identifies transform revision being used. \\
    fk (Flow Execution) & flow\_execution\_id & Identifies flow execution this execution is recorded as a part of. \\
    \bottomrule
  \end{tabular}
  \caption{Transform Execution reference.}
\end{table} 

\subsection{Transform Execution Slot (Entity)}
\label{sec:ref_transform_execution_slot}

Instance of transform slots populated with an additional runtime properties. Dataset revision affiliation is established and recorded at this level. 

\begin{table}[!htbp]
  \centering
  \begin{tabular}{p{0.25\linewidth}p{0.25\linewidth}p{0.4\linewidth}}
    \toprule
    Key & Name & Description \\
    \midrule
    fk (Transform Execution) & transform\_execution\_id & Identifies transform execution. \\
    fk (Transform Slot) & transform\_slot\_id & Identifies transform revision slot. \\
    fk (Dataset) & dataset\_id & Identifies dataset transform execution interacted with. \\
    fk (Dataset Revision) & dataset\_revision\_id & Identifies dataset revision (produced or consumed). \\
    \bottomrule
  \end{tabular}
  \caption{Transform Execution Slot reference.}
\end{table} 

\subsection{Flow (Entity)}
\label{sec:ref_flow}

Defines computational graph consisting of individual transforms. 

\begin{table}[!htbp]
  \centering
  \begin{tabular}{p{0.25\linewidth}p{0.2\linewidth}p{0.4\linewidth}}
    \toprule
    Key & Name & Description \\
    \midrule
    ek (External Provider) & external\_provider\_id & Identifies external flow definition system (e.g. AWS Step Functions, Apache Airflow). \\
    ek (External Flow) & external\_flow\_id & Flow identifier in the external system. \\
    & name & Flow name. \\
    \bottomrule
  \end{tabular}
  \caption{Flow reference.}
\end{table} 

\subsection{Flow Revision (Entity)}
\label{sec:ref_flow_revision}

Refers to the specific revision of the flow. Revisions store flow definitions and are used to keep track of changes occurring within those. 

\begin{table}[!htbp]
  \centering
  \begin{tabular}{p{0.25\linewidth}p{0.2\linewidth}p{0.4\linewidth}}
    \toprule
    Key & Name & Description \\
    \midrule
    ek (External Revision) & external\_revision\_id & Identifies parent flow. \\
    & definition & Flow definition is provided for completeness to incorporate proper mapping between internal and external transform identifiers. \\
    \bottomrule
  \end{tabular}
  \caption{Flow Revision reference.}
\end{table} 

\subsection{Flow Execution (Entity)}
\label{sec:ref_flow_execution}

Refers to the specific execution of the flow revision. 

\begin{table}[!htbp]
  \centering
  \begin{tabular}{p{0.25\linewidth}p{0.2\linewidth}p{0.4\linewidth}}
    \toprule
    Key & Name & Description \\
    \midrule
    fk (Flow Revision) & flow\_revision\_id & Identifies flow revision being invoked. \\
    & definition & Contains overview of the execution log with references to involved transform executions. \\
    \bottomrule
  \end{tabular}
  \caption{Flow Execution reference.}
\end{table} 

\subsection{Group (Entity)}
\label{sec:ref_group}

Group concept is layered on top of the entire model and can be used to relate a set of other entities that logically belong together. 

\begin{table}[!htbp]
  \centering
  \begin{tabular}{p{0.2\linewidth}p{0.2\linewidth}p{0.5\linewidth}}
    \toprule
    Key & Name & Description \\
    \midrule
    fk L(Entity) & items & The list of items marked for inclusion in the group. \\
    \bottomrule
  \end{tabular}
  \caption{Group reference.}
\end{table}  

\subsection{Identity (Entity)}
\label{sec:ref_identity}

Identity represents actor being tracked in the external identity management system. 

\begin{table}[!htbp]
  \centering
  \begin{tabular}{p{0.25\linewidth}p{0.2\linewidth}p{0.4\linewidth}}
    \toprule
    Key & Name & Description \\
    \midrule
    ek (External  Provider) & external\_provider\_id & Identifier of external identity provider. \\
    ek (External Actor) & external\_actor\_id & Identifier of actor within provider space. \\
    \bottomrule
  \end{tabular}
  \caption{Identity reference.}
\end{table} 

\subsection{Transaction (Entity)}
\label{sec:ref_transaction}

Transactions are used for recording lineage tracking model changes. 

\begin{table}[!htbp]
  \centering
  \begin{tabular}{p{0.2\linewidth}p{0.2\linewidth}p{0.5\linewidth}}
    \toprule
    Key & Name & Description \\
    \midrule
    fk (Identity) & identity\_id & Refers to identity used to execute transaction. \\
    & timestamp & Moment in time when transaction is committed. \\
    & additions & The list of additions introduced by the transaction. \\
    & modifications & The list of modifications introduced by the transaction. \\
    & removals & The list of removals introduced by the transaction. \\
    \bottomrule
  \end{tabular}
  \caption{Transaction reference.}
\end{table}

\end{document}